\documentclass[nofootinbib,aps,amssymb,floatfix,superscriptaddress,preprintnumbers]{revtex4}

\usepackage{textcomp}
\usepackage[latin1]{inputenc}                    
\usepackage{graphicx}                            
\usepackage{epsfig}
\usepackage[export]{adjustbox}
\usepackage{mathrsfs}
\usepackage{mathtools}
\usepackage{tabularx}
\usepackage{float}
\usepackage{enumitem}

\usepackage{epsf}
\usepackage{latexsym}                            
\usepackage{amsfonts}                            
\usepackage{amssymb}                             
\usepackage{amsmath}                             
\usepackage{slashed}
\usepackage[mathscr]{eucal}                      
\usepackage{dcolumn}                             
\usepackage{multirow}                               
\usepackage{bm}                                  
\usepackage{hyperref}                            
\usepackage[usenames]{color}
\usepackage[dvipsnames]{xcolor}

\usepackage{color}
\usepackage[normalem]{ulem}

\begin{document}

\title{First global QCD analysis of the TMD $\boldsymbol{g_{1T}}$ from semi-inclusive DIS data}

\author{Shohini Bhattacharya}
\affiliation{Physics Department, Brookhaven National Laboratory, Building 510A, Upton, NY 11973, USA}
\affiliation{Department of Physics,  Temple University,  Philadelphia,  PA 19122, USA}
\author{Zhong-Bo Kang}
\affiliation{Department of Physics and Astronomy, University of California, Los Angeles, CA 90095, USA}
\affiliation{Mani L. Bhaumik Institute for Theoretical Physics,
University of California, Los Angeles, CA 90095, USA}
\affiliation{Center for Frontiers in Nuclear Science, Stony Brook University, Stony Brook, NY 11794, USA}
\author{Andreas Metz}
\affiliation{Department of Physics,  Temple University,  Philadelphia,  PA 19122, USA}
\author{Gregory Penn}
\affiliation{Department of Physics, Yale University, New Haven, CT 06511, USA}
\affiliation{Department of Physics,  Temple University,  Philadelphia,  PA 19122, USA}
\author{Daniel Pitonyak}
\affiliation{Department of Physics, Lebanon Valley College, Annville, PA 17003, USA}

\begin{abstract}
\noindent The worm-gear transverse momentum dependent (TMD) function $g_{1T}$ is one of the eight leading-twist TMDs and has the probabilistic interpretation of finding a longitudinally polarized quark inside a transversely polarized hadron. In this work, we present the first simultaneous extraction of $g_{1T}$ from all the available experimental measurements. The study involves the analysis of COMPASS, HERMES, and Jefferson Lab data on semi-inclusive deep-inelastic scattering using Monte Carlo techniques. We also compare $g_{1T}$ obtained from this experimental data with different theoretical approaches, such as the large-$N_c$ approximation, the Wandzura-Wilczek-type approximation, and lattice QCD.  
\end{abstract}

\maketitle

\section{Introduction}
\label{sec:introduction}
An important class of functions required to understand the structure of hadrons in terms of their underlying partons are transverse momentum dependent parton distribution and fragmentation functions (TMD PDFs and TMD FFs, collectively called TMDs). We use TMDs in the description of hard scattering processes such as semi-inclusive deep-inelastic scattering (SIDIS), Drell-Yan, and $e^{+} \, e^{-}$ annihilation into hadron pairs. 
The presence of both hard and soft scales in these reactions allows one to probe the intrinsic motion of partons.  For example, in SIDIS one has the photon virtuality $q^2\equiv -Q^2$ and transverse momentum $q_T$ such that if $q_T \ll Q$, one is sensitive to TMD physics. While the usual collinear PDFs describe the probability to find a parton in a fast-moving hadron with a particular fraction $x$ of the hadron's longitudinal momentum, TMD PDFs in addition encode the probability that the parton has a specific
transverse momentum $\vec{k}_\perp$. The TMD PDFs can therefore be regarded as the natural extension of collinear PDFs from one to three dimensions in momentum space.

When the spin of the hadron and that of the quarks are taken into account, certain constraints of QCD, such as Hermiticity, time reversal, and parity, allow us to define eight TMDs at leading twist (twist-2) as functions of $(x, \vec{k}^{2}_\perp)$: $f_{1}$ (unpolarized function), $g_{1L}$ (helicity function), $h_{1}$ (transversity function), $f^{\perp}_{1T}$ (Sivers function), $g_{1T}$ and $h^{\perp}_{1L}$ (``worm-gear'' functions), $h^{\perp}_{1}$ (Boer-Mulders function) and $h^{\perp}_{1T}$ (pretzelosity function).
Only three of them, $f_{1}$, $g_{1L}$, and $h_{1}$, survive after carrying out the $\vec{k}_\perp$ integral, while the remaining five vanish. Those
five TMDs provide novel information about spin-orbit correlations. Just like the collinear PDFs, extracting TMDs calls for global fits of experimental data.
There has been tremendous progress in our understanding of TMDs and their extraction from data in the last few years, thanks to new theoretical/phenomenological ideas and experimental measurements at, e.g, Belle, COMPASS, HERMES, Jefferson Lab (JLab), and the Relativistic Heavy Ion Collider (RHIC) at Brookhaven National Lab (BNL). From this perspective, one of the least known TMDs is $g_{1T}$, which is the focus of this work. This function has the probabilistic interpretation of finding a longitudinally polarized quark inside a transversely polarized hadron~\cite{Tangerman:1994eh,Kotzinian:1994dv,Mulders:1995dh,Boer:1997nt,Goeke:2005hb}. Since the quark ``spins'' in one direction while the hadron ``spins'' perpendicular to that, it is also known as a ``worm-gear" function.

There does exist some prior information on $g_{1T}$ from various sources. Quite a few model calculations for  the function exist, e.g., in the light-cone constituent quark model~\cite{Bacchetta:1999kz,Ji:2002xn,Pasquini:2008ax,Boffi:2009sh,Bacchetta:2010si}, spectator diquark model~\cite{Jakob:1997wg,Gamberg:2007wm,Bacchetta:2008af,Bacchetta:2010si}, MIT bag model~\cite{Avakian:2010br}, and the covariant parton model~\cite{Efremov:2009ze}.  All of these calculations suggest that the up quark function $g^{u}_{1T}$ is positive, down quark function $g^{d}_{1T}$ is negative, and the peak amplitude for $g^{u}_{1T}$ is larger in magnitude than that of $g^{d}_{1T}$.  In addition, the first lattice QCD calculations of TMDs were presented in Refs.~\cite{Hagler:2009mb,Musch:2010ka,Yoon:2017qzo}, which demonstrated the $k_\perp$-moment of $g_{1T}(x, \vec{k}^{2}_\perp)$, denoted $g^{(1)}_{1T}(x)$, gives rise to distortions in the quark densities in the transverse momentum plane. The size of the distortion was characterized by a quantity called the ``transverse momentum shift"~$\langle \rho_{x} \rangle ^q_{TL}$ for a quark flavor~$q$:
$
\langle \rho_{x} \rangle ^q_{TL} \sim \int^{1}_{0} dx \, \big ( g^{(1)q}_{1T}(x) - g^{(1)\bar{q}}_{1T}(x) \big ),
$
where $g^{(1)\bar{q}}_{1T}(x)$ is the antiquark distribution.  This shift was found to be positive for up quarks and negative for down quarks. Furthermore, this study determined that the amount of shift was larger for the up quarks.  These lattice results certainly indicate that $g^{u}_{1T}$ and $g^{d}_{1T}$ differ in sign and possibly also have relatively different magnitudes. Finally, there are two existing theoretical predictions for $g_{1T}$ based on certain approximations.  The large-$N_c$ approximation~\cite{Pobylitsa:2003ty} states that $g^{u}_{1T}=-g^{d}_{1T}$ up to $1/N_c$ corrections.  The Wandzura-Wilczek (WW)-type approximation~\cite{Avakian:2007mv,Accardi:2009au,Kanazawa:2015ajw,Scimemi:2018mmi} neglects quark-gluon-quark correlators to arrive at $g_{1T}^{(1)}(x) = x\int_x^1dy\,g_1(y)/y$, where $g_1$ is the helicity PDF.  The WW-type approximation has been used to make predictions for the relevant asymmetries in SIDIS~\cite{Kotzinian:2006dw,Avakian:2007mv,Bastami:2018xqd, Benic:2021gya}, single-inclusive (collinear twist-3) reactions in lepton-nucleon collisions~\cite{Kang:2011jw,Kanazawa:2014tda}, and vector boson production from proton-proton collisions~\cite{Huang:2015vpy}. Still, to date, there has been no proper extraction of $g_{1T}$ from experimental data.

In this work, we perform for the first time, using Monte Carlo (MC) techniques, a global QCD extraction of the TMD $g_{1T}$ from all available data, namely, SIDIS measurements from HERMES~\cite{HERMES:2020ifk}, COMPASS~\cite{COMPASS:2016led,Parsamyan:2018evv,Avakian:2019drf}, and JLab~\cite{JeffersonLabHallA:2011vwy}. 
We also provide a quantitative comparison of $g_{1T}$ obtained from experimental data to the large-$N_c$ approximation, the WW-type approximation, and lattice QCD calculations. The work presented in this paper is an important step toward fully understanding the transverse (partonic) structure of the nucleon. 
We organize the manuscript as follows. In Sec.~\ref{sec:theory}, we provide the definition of $g_{1T}$ and discuss the large-$N_c$ and the WW-type approximations. We also provide a brief sketch of how SIDIS acts as a probe for $g_{1T}$. In Sec.~\ref{sec:MC_technqiue}, we discuss the parameterization for $g_{1T}$ and give an overview of the MC techniques that we use in performing the fit. In Sec.~\ref{sec:fits}, we present our results and also make the aforementioned comparisons to other information we have on $g_{1T}$. In Sec.~\ref{sec:summary}, we summarize our work and provide a brief outlook.

\section{Theoretical Background}
\label{sec:theory}
The TMD $g_{1T}(x,\vec{k}_\perp^2)$, for a quark with momentum $k$, is defined as the following projection of the quark-quark correlator for a transversely polarized nucleon with momentum $P$ and spin $S_\perp=(0^+,0^-,\vec{S}_\perp)$~\cite{Tangerman:1994eh,Kotzinian:1994dv,Mulders:1995dh,Boer:1997nt,Goeke:2005hb},
\begin{align}
\Phi^{[\gamma^{+} \gamma_{5}]}(x, \vec{k}_\perp, \vec{S}_\perp) & =\frac{1}{2} \int \frac{dz^{-}d^{2}\vec{z}_\perp}{(2\pi)^{3}} e^{i k \cdot z} \langle P,S_{\perp}| \bar{\psi} (0)\gamma^{+}\gamma_{5} \, {\cal W}(0,z) \psi (z)|P,S_{\perp} \rangle \bigg |_{z^{+}=0} \nonumber \\[0.2cm]
& = \dfrac{\vec{k}_{\perp}\cdot\vec{S}_{\perp}}{M} g_{1T}(x, \vec{k}^{2}_\perp) \, ,
\label{e:TMD_correlator}
\end{align}
where $x\equiv k^+/P^+$, $M$ is the nucleon mass, and ${\cal W}(0,z)$ is a Wilson line connecting the quark fields that ensures color gauge invariance.  
The main outcome of this paper is the first global fit of $g_{1T}$ from experimental data.  There is already some information about $g_{1T}$ from certain approximations based on theoretical analyses:
\begin{itemize}
\item The large-$N_{c}$ approximation states that, in the limit of a large number of colors $N_c$, $g_{1T}$ for up and down quarks are related as~\cite{Pobylitsa:2003ty},
\begin{align}
g^{u}_{1T} (x, \vec{k}^{2}_\perp) & \stackrel{\text{Large-$N_c$}}{\approx} -g^{d}_{1T} (x, \vec{k}^{2}_\perp) \, ,
\label{e:largeNc_g1T}
\end{align}
where terms of relative $\mathcal{O}(1/N_c)$ have been neglected.
\item The first $\vec{k}_\perp$-moment $g_{1T}^{(1)}(x)$ can be related to an integral of the helicity PDF $g_{1}(x)$ and a term involving quark-gluon-quark correlators~\cite{Avakian:2007mv,Accardi:2009au,Kanazawa:2015ajw,Scimemi:2018mmi}.  The Wandzura-Wilczek (WW)-type approximation neglects the latter to arrive at
\begin{align}
g^{(1)}_{1T}(x) & \equiv \int d^{2}\vec{k}_\perp \bigg ( \dfrac{k^{2}_\perp}{2M^{2}}\bigg ) g_{1T}(x, \vec{k}^{2}_\perp) \stackrel{\text{WW-type}}{\approx} x \int^{1}_{x} \dfrac{dy}{y} g_{1}(y) \, .\label{e:WWtype}
\end{align}
\end{itemize}
In Sec.~\ref{sec:fits}, we will use our global fitting results from experimental data to test for the first time these theoretical relationships  as well the calculations from lattice QCD~\cite{Hagler:2009mb,Musch:2010ka,Yoon:2017qzo}.

The SIDIS process is an excellent channel to probe the transverse structure of the nucleon. We can write the reaction as
\begin{align}
\ell  (l,\lambda_{l}) \, + \, N(P,S) \rightarrow \ell' (l',\lambda'_{l}) \, + \, h(P_{h}) \, + \, X \, ,
\end{align}
where, $\ell  (l,\lambda_{l})$ ($\ell' (l',\lambda'_{l})$) denotes an incoming (outgoing) lepton, $N(P,S)$ denotes a nucleon, and $h(P_{h})$ denotes a measured hadron in the final state, with other unobserved particles denoted by the symbol $X$. The momenta and polarizations of the particles involved are indicated in parentheses.
There are six independent kinematical variables for this process,
\begin{align}
x_B = \dfrac{Q^{2}}{2P \cdot q} \,, \quad Q^{2} = -(l-l')^2 \, , \quad \phi_{S} \, , \quad z_{h} = \dfrac{P \cdot P_{h}}{P \cdot q} \, , \quad P_{hT} \,, \quad \phi_{h} \, .
\end{align}
In the one-photon exchange approximation, $Q^{2}$ is the virtuality of the photon.  We work at leading order, where $x_B$ ($z_h$) is equivalent to the fraction $x$ ($z$) of the incoming nucleon's (fragmenting quark's) momentum carried by the struck quark (produced hadron), so we drop the subscripts for brevity. In the photon-nucleon center-of-mass frame, the azimuthal angle $\phi_{S}$ gives the orientation of the (transverse) spin vector of the nucleon $\vec{S}_\perp$ relative to the lepton plane (plane formed by the incoming and outgoing lepton),  $P_{hT}=(0^+,0^-,\vec{P}_{hT})$ is the transverse momentum of the produced hadron, and $\phi_{h}$ gives the orientation of $\vec{P}_{hT}$ relative to the lepton plane.  One can also work in a reference frame where the hadron has no transverse momentum.  In that case, the virtual photon has transverse momentum $q_T$, and one has $q_T = P_{hT}/z$ up to $\mathcal{O}(1/Q^2)$ corrections.  The factorization of this process in terms of TMDs requires that $q_T \ll Q$. One can provide a model-independent decomposition of the (one-photon exchange) SIDIS cross section in terms of a certain number of structure functions $F_{XY}$, with $X$ and $Y$ denoting the polarization (unpolarized ($U$), longitudinal ($L$), or transverse ($T$)) of the incoming lepton and nucleon, respectively. The (six-fold) differential cross section reads~\cite{Bacchetta:2006tn,Bastami:2018xqd}, 
\begin{align}
\dfrac{d\sigma}{dx\, dy \, d\phi_{S}\, dz_{h} \, d \phi_{h} \, d P^{2}_{hT}} & = \dfrac{\alpha^{2}_{\rm em}}{x \, y \, Q^{2}} \bigg \{ \bigg ( 1-y+\dfrac{1}{2} y^{2} \bigg ) F_{UU}  \nonumber \\[0.2cm]
&  + \lambda_{l} \, |\vec{S}_{\perp}| \, y \bigg ( 1 - \dfrac{1}{2}y \bigg ) \, {\rm cos} (\phi_{h}-\phi_{S}) F^{{\rm cos}(\phi_{h}-\phi_{S})}_{LT} + \dots \bigg \} \, ,
\label{e:SIDIS_cross_section}
\end{align}
where $\alpha_{\rm em}$ is the fine structure constant, $y = P \cdot q / P \cdot l$, and $\dots$ denotes terms that are irrelevant for our study. The structure functions $F_{XY}$ depend on $(x, z, P^{2}_{hT}, Q^{2})$, which we did not write explicitly in the above expressions. One can separately isolate each structure function since each has its own unique angular modulation associated with it.

In the limit $q_T \ll Q$, we can use TMD factorization~\cite{Collins:1981uk,Collins:1984kg,Meng:1995yn,Ji:2004xq,Collins:2011zzd} to write the cross section in terms of perturbatively calculable hard interactions and (non-perturbative) TMDs. The resulting expressions for the structure functions then involve convolutions of the TMD PDFs and FFs (generically denoted $f(x,\vec{k}_\perp^2)$ and $D(z,\vec{P}_\perp^2)$, respectively)~\cite{Bacchetta:2006tn,Bastami:2018xqd}, 
\begin{align}
C \Big [w \, f \, D \Big ] = 
x \sum_{q} e^{2}_{q} \int d^{2} \vec{k}_{\perp} \int d^{2}\vec{P}_{\perp} \,  \delta^{(2)} \big ( z \vec{k}_{\perp} + \vec{P}_{\perp} - \vec{P}_{hT} \big ) \, w (\vec{k}_\perp, \vec{P}_\perp) \, f^{q} (x, \vec{k}^{2}_{\perp}) \, D^{q} (z, \vec{P}^{2}_{\perp}) \, ,
\label{e:convolution_g1T}
\end{align}
where $w$ is a weight factor depending on the transverse momenta $\vec{k}_\perp,\vec{P}_\perp$. The structure functions in Eq.~(\ref{e:SIDIS_cross_section}) read~\cite{Bacchetta:2006tn,Bastami:2018xqd}, 
\begin{align}
F_{UU} &= C \Big [ w^{\{0\}} \, f_{1} (x, \vec{k}^{2}_{\perp}) \, D_{1} (z, \vec{P}^{2}_{\perp} ) \Big ] \, ,
\label{e:FUU}
\\[0.2cm]
F^{{\rm cos} (\phi_{h}-\phi_{S})}_{LT} &= C \Big [ w^{\{1\}} \, g_{1T} (x, \vec{k}^{2}_{\perp}) \, D_{1} (z, \vec{P}^{2}_{\perp} ) \Big ] \, ,
\label{e:FLT}
\end{align}
where $f_1,D_{1}$ are the unpolarized TMDs, and 
$w^{\{0\}} = 1 \, , w^{\{1\}} = \vec{P}_{hT}\cdot \vec{k}_{\perp}/(|\vec{P}_{hT}| M)$.
We will work with the leading order (parton model) expression for the cross section and a Gaussian ansatz for the TMDs,
\begin{align}
g^{q}_{1T} (x, \vec{k}^{2}_{\perp}) &= g^{(1)q}_{1T}(x) \dfrac{2 M^{2} e^{-\tfrac{\vec{k}^{2}_{\perp}}{\langle k^{2}_{\perp} \rangle |_{g_{1T}^q}}}}{\pi \big ( \langle k^{2}_{\perp} \rangle |_{g_{1T}^q} \big )^2 }  \, , \quad  
f^{q}_{1} (x, \vec{k}^{2}_{\perp}) = f^{q}_{1}(x) \dfrac{e^{-\tfrac{\vec{k}^{2}_{\perp}}{\langle k^{2}_{\perp} \rangle |_{f_1^q}}}}{\pi \langle k^{2}_{\perp} \rangle |_{f_1^q} }  \, , 
\quad
D^{q}_{1} (z, \vec{P}^{2}_{\perp}) = D^{q}_{1}(z) \dfrac{e^{-\tfrac{\vec{P}^{2}_{\perp}}{\langle P^{2}_{\perp} \rangle_q }}}{\pi \langle P^{2}_{\perp} \rangle_q} \, , \label{e:Gauss}
\end{align}
which has been quite successful in describing a wide variety of reactions~\cite{Anselmino:2005nn,Anselmino:2000vs,Anselmino:2005ea,Vogelsang:2005cs,Collins:2005ie,Collins:2005rq,Anselmino:2007fs,Anselmino:2008jk,Schweitzer:2010tt,Qiu:2011ai,Anselmino:2013vqa,Signori:2013mda,Anselmino:2013lza,Boer:2014tka,DAlesio:2020wjq,Callos:2020qtu,Cammarota:2020qcw} and is sufficient to gain the first information on $g_{1T}$ from experimental data. In Eq.~(\ref{e:Gauss}), $\langle k^{2}_{\perp} \rangle$ and $\langle P^{2}_{\perp} \rangle$ are the TMD PDF and the FF widths, which can in principle be flavor dependent.  Also, lattice QCD data are compatible with the Gaussian shape for the TMDs at small transverse momenta~\cite{Hagler:2009mb, Musch:2010ka}.
Using these, we obtain the following expressions for the structure functions, 
\begin{align}
F_{UU} & = x \, \sum_{q} \, e^{2}_{q} \, f^{q}_{1} (x) \, D^{q}_{1}(z) \, g^{q}_{f_1} \, , \\
F_{LT}^{{\rm cos} (\phi_{h}-\phi_{S})} & = x \, \sum_{q} \, e^{2}_{q}\, g^{(1)q}_{1T} (x) D^{q}_{1} (z) \, (2M) \, (z \, P_{hT}) \, \dfrac{g^{q}_{g_{1T}}}{\lambda^{q}_{g_{1T}}} \, ,
\end{align}
with
\begin{align}
\lambda^{q}_{f} = z^{2} \langle k^{2}_{\perp} \rangle \big |_{f^{q}} \, + \, \langle P^{2}_{\perp} \rangle \big |_{D^{q}_{1}} \, , \qquad
g^{q}_{f} = \dfrac{e^{-P^{2}_{hT} /\lambda^{q}_{f}}}{\pi \lambda^{q}_{f}}  \, ,
\end{align}
where $f = g_{1T}\;{\rm or}\; f_{1}$. These results agree with Refs.~\cite{Mulders:1995dh,Bastami:2018xqd}. We use them to fit $g^{(1)}_{1T}(x)$ and $\langle k^{2}_{\perp} \rangle \big |_{g_{1T}^{q}}$ to the experimental data for the asymmetry $A^{{\rm cos} (\phi_{h}-\phi_{S})}_{LT}$, which is defined as
\begin{align}
A^{{\rm cos} (\phi_{h}-\phi_{S})}_{LT} & = \dfrac{F^{{\rm cos} (\phi_{h}-\phi_{S})}_{LT}}{F_{UU}} \, .
\end{align}

\section{Overview of the Fitting Methodology}
\label{sec:MC_technqiue}
As put forth in Eq.~(\ref{e:Gauss}), we use a Gaussian function for the $k_\perp$-dependence of $g_{1T}(x,\vec{k}_\perp^2)$, with $\langle k^{2}_{\perp} \rangle \big |_{g_{1T}^{q}}$ a free parameter, while the $x$-dependence is encoded in its first $k_\perp$-moment $g_{1T}^{(1)}(x)$. We parameterize this latter function, which also depends on $Q^2$, as
\begin{align}
g_{1T}^{(1)} (x, Q^{2}) & = \dfrac{N}{\widetilde{N}} x^{\alpha} (1-x)^{\beta} f_{1} (x, Q^{2}) \, ,
\label{e:g1T_para}
\end{align}
where $N$ is a normalization parameter, $\alpha$ and $\beta$ are parameters that determine the behavior of $g^{(1)}_{1T}(x)$ as $x\rightarrow 0$ and $x \rightarrow 1$, respectively, and $f_1(x,Q^2)$ is the unpolarized PDF.  We normalize the function so that, at the initial scale $Q_{0}\equiv 2\, {\rm GeV}$,
$
\int^{1}_{0} dx \, x \, g^{(1)}_{1T} (x, Q^{2}_{0}) = N.
$
That is, 
$\widetilde{N} \equiv \int_{0}^{1} dx \, x^{\alpha+1} (1-x)^{\beta} f_{1}(x, Q^{2}_{0}).$
We take the evolution of $g_{1T}^{(1)}(x)$ to be the same as the DGLAP evolution of $f_1(x)$ and use CT10 PDFs~\cite{Lai:2010vv} at next-to-leading order for the latter. The same unpolarized PDF set is used in calculating $F_{UU}$.  Even though $g_{1T}^{(1)}(x)$ follows a more complicated evolution that also mixes with quark-gluon-quark correlators, the ``diagonal'' piece does contain a term that is the same as the unpolarized DGLAP splitting function~\cite{Zhou:2008mz}.  A similar structure occurs in the evolution of the Qiu-Sterman function~\cite{Braun:2009mi,Kang:2012em,Schafer:2012ra} (relevant for the Sivers effect), and Ref.~\cite{Echevarria:2020hpy} found the fit of SIDIS data does not significantly change if one includes more terms in the evolution than just unpolarized DGLAP.  Therefore, the evolution of $g_{1T}^{(1)}(x)$ implicit in Eq.~(\ref{e:g1T_para}) is a reasonable first approximation.  We also explicitly checked that using different unpolarized PDF sets in our fit (like CT18~\cite{Hou:2019efy} next-to-leading order and CJ15~\cite{Accardi:2016qay} leading order) causes very minimal changes to our extracted $g_{1T}^{(1)}(x)$.

\begin{figure}[t]
\centering
\includegraphics[width=0.5\textwidth]{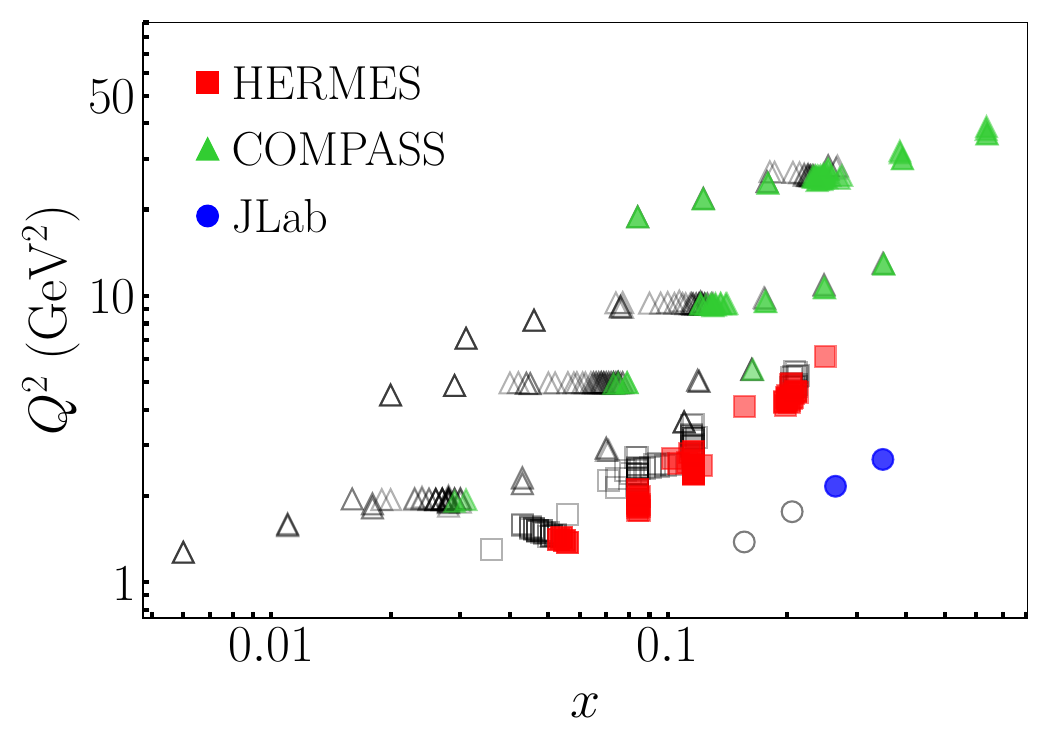}
\caption{The $x$-$Q^2$ coverage for the HERMES~\cite{HERMES:2020ifk} (squares), COMPASS~\cite{COMPASS:2016led} (triangles), and JLab~\cite{JeffersonLabHallA:2011vwy} (circles) data sets. Colored (open) points are for data that do (do not) satisfy the cut of $q_{T}/Q < 0.50$.}
\label{f:Q2_vs_x}
\end{figure}

Our ansatz for the functional form and evolution of $g_{1T}(x,\vec{k}_\perp^2)$ are sufficient to work with at the present stage since the available data is not precise enough, nor spans a large enough range in $Q^2$, to be sensitive to the finer details of TMD evolution or a more flexible functional form.  If data on $A_{LT}$ from vector boson production in proton-proton collisions becomes available in the future, such issues will need to be revisited. 
Note that for simplicity of notation, we have not shown the explicit flavor dependence of the parameters in Eq.~(\ref{e:g1T_para}). For the actual fit, we do assign flavor dependence to most of the parameters, as we discuss below.
We also assume $g^{(1)\bar{u}}_{1T}(x)=g^{(1)\bar{d}}_{1T}(x)=g^{(1)s}_{1T}(x)=g^{(1)\bar{s}}_{1T}(x)=0$ since most of the data included in our fit is in the moderate-$x$ region (see Fig.~\ref{f:Q2_vs_x}). Note that one can also choose to work with the helicity PDF in Eq.~(\ref{e:g1T_para}) rather than the unpolarized PDF. We have explicitly confirmed, using $g_{1}(x)$ from Ref.~\cite{Ethier:2017zbq}, that the extracted $g_{1T}^{(1)}(x)$ remains essentially unchanged.

We start with seven free parameters: $\langle k^{2}_\perp \rangle\big |_{g_{1T}^u} = \langle k^{2}_\perp \rangle\big |_{g_{1T}^d}\equiv \langle k^{2}_\perp \rangle\big |_{g_{1T}}$ and $N^{q}$, $\alpha^{q}$, $\beta^{q}$ for $q=u,d$. However, we find that the presently available data is insufficient to constrain $\langle k^{2}_\perp \rangle\big |_{g_{1T}}$, $\alpha^d$, $\beta^{u},\beta^d$. Therefore, for the final fit we work with three free parameters:~$N^{u},N^d$, and $\alpha^{u}=\alpha^{d}\equiv \alpha$. As for $\langle k^{2}_\perp \rangle\big |_{g_{1T}}$, we fix it according to what is supported by the lattice QCD calculation of Ref.~\cite{Hagler:2009mb} involving the widths of the unpolarized and helicity PDFs ($\langle k^{2}_\perp \rangle \big |_{f_{1}}$ and $\langle k^{2}_\perp \rangle \big |_{g_{1}}$, respectively),
\begin{align}
\dfrac{\langle k^{2}_\perp \rangle \big |_{g_{1}}}{\langle k^{2}_\perp \rangle \big |_{f_{1}}} \approx \dfrac{{\langle k^{2}_\perp \rangle \big |_{g_{1T}}}}{\langle k^{2}_\perp \rangle \big |_{f_{1}}} \approx 0.76 \, ,
\label{e:TMD_width_lattice}
\end{align}
where we have assumed $\langle k^{2}_\perp \rangle \big |_{g_{1}}\approx \langle k^{2}_\perp \rangle \big |_{g_{1T}}$.  We note the approximation (\ref{e:TMD_width_lattice}) was first used in Ref.~\cite{Bastami:2018xqd}. In our case, since we take $\langle k^{2}_\perp \rangle \big |_{f_1} = 0.53$ $\rm{GeV}^{2}$ from Ref.~\cite{Cammarota:2020qcw}, this leads to $\langle k^{2}_\perp \rangle \big |_{g_{1T}} = 0.40 \, \rm{GeV}^{2}$.   
As for $\beta^u,\beta^d$, HERMES and JLab do not provide measurements at very large $x$. On the other hand, COMPASS does, but the errors increase significantly in that region. Therefore, because of the lack of precise data at large $x$, we are unable to constrain the $\beta$ parameter and instead fix it according to the WW-type approximation (\ref{e:WWtype}):~since $g_1(x)\sim (1-x)^{\beta_{g_1}}$ at large $x$, the WW-type approximation makes it reasonable to assume that $g^{(1)}_{1T}(x)$ falls off as $(1-x)^{\beta_{g_1}+1}$, and because $\beta_{g_1}\approx \beta_{f_1}$, we set $\beta^u=\beta^d  \equiv \beta =1$. We have also explored different values of $\langle k^{2}_\perp \rangle \big |_{g_{1T}}$ and $\beta$ within the ranges $0.1 \, \textrm{GeV}^{2}  < \langle k^{2}_\perp \rangle \big |_{g_{1T}} < \langle k^{2}_\perp \rangle \big |_{f_1}$ and $0 < \beta < 6$. 
(Note that for $\langle k^{2}_\perp \rangle \big |_{g_{1T}}$, it is important to stay within the {mentioned} range in order to be consistent with the positivity bound~\cite{Bacchetta:1999kz},
$
{k_\perp}/{M} \, \big| g_{1T}(x, \vec{k}^{2}_\perp) \big| <  f_{1}(x, \vec{k}^{2}_\perp) \, .)
$
We checked explicitly that none of the qualitative conclusions of our work change by choosing different values for $\langle k^{2}_\perp \rangle \big |_{g_{1T}}$ and $\beta$. Certainly more precise data will be required to better constrain the values of these parameters in the future.

We fit the experimental data and 200 additional replicas of the data, where each ``pseudo-data'' point $P_j$ in the replica is given by $P_j = D_j + R_j\cdot e_j$, where $D_j$ is the actual experimental data point, $R_j$ is a random number generated from a Gaussian distribution centered at $0$ with standard deviation of $1$, and $e_j$ is the quadrature sum of statistical and systematic experimental errors.  The starting parameters for the fit are chosen from a flat sampling of the parameter space and then Monte Carlo techniques are used to calculate the functions and observables from the posterior distributions. Namely, the average value $E[\mathcal{O}]$, used to create a central curve for a function or observable $\mathcal{O}$, is determined by $E[\mathcal{O}] = \frac{1} {n}\sum_i \mathcal{O}(\vec{p}_i)$, where $\vec{p}_i=\{N^u,N^d,\alpha\}_i$ is the set of parameters that minimizes the $\chi^2$ for replica $i$, with $n$ being the total number of replicas.  The 1-$\sigma$ confidence level (CL) error band is found from the standard deviation of $\mathcal{O}$:~$S[\mathcal{O}] = \sqrt{\frac{1} {n}\sum_i \left(\mathcal{O}(\vec{p}_i)-E[\mathcal{O}]\right)^2}$. 

\section{Phenomenological Results}
\label{sec:fits}
Before presenting our final fit results, we provide an overview of the experimental data. A set of 152 data points were reported by the HERMES Collaboration using a proton target with a pion ($\pi^{\pm}$ or $\pi^0$) detected in the final state~\cite{HERMES:2020ifk}. While the asymmetries for the neutral pion were extracted in one-dimensional kinematical bins, the ones for the charged pions were extracted (for the first time) in three-dimensional kinematical bins of $(x, z, P_{hT})$. A set of 396 data points were reported by the COMPASS Collaboration for a proton target with an unidentified charged hadron ($h^{\pm}$) in the final state~\cite{COMPASS:2016led,Parsamyan:2018evv,Avakian:2019drf}. These data were extracted in two-dimensional kinematical bins of $(x, Q^{2})$, $(z, Q^{2})$ and $(P_{hT}, Q^{2})$. Furthermore, these data were sub-divided into three $z$ intervals: $z>0.1$, $0.1<z<0.2$, and $z>0.2$. Here we choose to work with data from the $z$ intervals $0.1<z<0.2$ and $z>0.2$ since we found that they allow us to better constrain $g_{1T}$ than the $z>0.1$ data set alone. A set of eight data points were reported by Jefferson Lab Hall A for a neutron target, and the measured hadrons were $\pi^{\pm}$~\cite{JeffersonLabHallA:2011vwy}. For the neutron, we apply isospin symmetry on $g_{1T}$ and $f_{1}$ for the up and down quark flavors ($u \leftrightarrow d, \bar{u} \leftrightarrow \bar{d}$), while leaving strange quarks ($s, \bar{s}$) unchanged.

In order to ensure the applicability of TMD factorization, one needs $q_T\ll Q$. In the literature, cuts $q_{T}/ Q \lesssim 0.25$ have been used recently~\cite{Scimemi:2017etj,Scimemi:2019cmh,Bacchetta:2019sam,Bury:2021sue}. In our case, with such a cut we lose a significant number of data points and are unable to sufficiently constrain all the parameters in our fit. Therefore, we apply a cut of $q_{T}/Q < 0.50$ to balance the condition for TMD factorization with the need to have enough data points to meaningfully extract $g_{1T}$. (This is similar to the approach in Ref.~\cite{Echevarria:2020hpy}, where a cut of $q_{T}/ Q < 0.75$ was used for a fit of the Sivers function.)
After applying this cut, we are left with 60 data points from HERMES, 64 data points from COMPASS, and 4 data points from JLab. In Fig.~\ref{f:Q2_vs_x}, we show the distribution in $x$ and $Q^{2}$ of the data points along with which ones survive our cut. 

As mentioned above, for COMPASS the measured hadrons are unidentified charged hadrons, and we make the approximation $D_1^{h^\pm} = D_1^{\pi^\pm} + D_1^{K^\pm}$. (All FFs are taken from DSS~\cite{deFlorian:2007ekg} at leading order.) We assign favored widths $\langle P^{2}_{\perp} \rangle \big |_{\rm{fav}}$ (unfavored widths $\langle P^{2}_{\perp} \rangle \big |_{\rm{unfav}}$) to $u$ and $\bar{d}$ ($\bar{u},d,s$, and $\bar{s}$) for asymmetries associated with $\pi^{+}$ production and employ charge conjugation for the $\pi^-$ FF; for $\pi^0$, we use $D_1^{\pi^0}=(D_1^{\pi^+}+D_1^{\pi^-})/2$. The explicit values for the widths of the FFs, which we take from Ref.~\cite{Cammarota:2020qcw}, are $\langle P^{2}_{\perp} \rangle \big |_{\rm{fav}}=0.124 \, {\rm GeV}^2$ and $\langle P^{2}_{\perp} \rangle \big |_{\rm{unfav}}=0.145 \, {\rm GeV}^{2}$.
We mention that HERMES does have data for $A_{LT}^{\cos(\phi_h-\phi_S)}$ for $K^\pm$.  However, since we are focused on extracting $g_{1T}$ for up and down quarks, and have set antiquarks and strange quarks to zero, we do not consider this data in our analysis.

\subsection{Main results:~Weighted $\chi^{2}$ method}
\label{sec:main_fit}
In this section, we present our final results for $g^{(1)}_{1T}(x)$ extracted simultaneously from HERMES, COMPASS, and JLab data in the so-called ``weighted $\chi^2$ method". The reason we utilize this approach, which we describe below in more detail, is that JLab has very few data points compared to HERMES and COMPASS. Therefore, in the process of fitting with the usual definition of $\chi^2$, 
\begin{equation}
\chi^2 = \sum_i\left(\frac{T_i - D_i}{e_i}\right)^{\!\! 2}\,, \label{e:chi2}
\end{equation}
where $T_i$ is the theory value for an experimental data point $D_i$ with error $e_i$, our description of the JLab data, specifically for $\pi^-$ (see Table~\ref{table:summary_chi2} in Sec.~\ref{sec:unweighted_chi2}), was not as good as the charged pion data sets from HERMES and COMPASS. (Observing such a feature probably is not surprising. Due to the larger number of data points from HERMES and COMPASS, the fit tries to find solutions to accommodate these data sets better and compromises on the quality of the fit to the JLab data where the number of data points is much less, and, consequently, so is their contribution to the overall $\chi^2$.) In an attempt to achieve an equally good description of all the experimental data sets, we weight the $\chi^2$ so as to emphasize the JLab data as much as the data from HERMES and COMPASS. This approach was recently used in Ref.~\cite{Echevarria:2020hpy} for a fit of the Sivers function.
Following Ref.~\cite{Echevarria:2020hpy}, we define a weighted $\chi^2$ ($\chi^2_w$) as
\begin{align}
\chi^{2}_w & = \chi^{2} \big |_{\rm {H+C}} + w \, \chi^{2} \big |_{\rm J} \, ,
\label{e:chi2_w}
\end{align}
where $\chi^{2} |_{\rm {H+C}}$ ($\chi^{2} |_{\rm J}$) is the $\chi^2$ for the HERMES and COMPASS (JLab) data points using the definition (\ref{e:chi2}). In Eq.~(\ref{e:chi2_w}), we choose the weight factor $w$ as $w \equiv N_{\rm H+C}/N_{\rm J} = 124/4$, where $N_{\rm H+C}$ ($N_{\rm J}$) is the total number of points from HERMES and COMPASS (JLab). We must also take
\begin{align}
N_{\rm pts.} & = N_{\rm H+C} + w \, N_{\rm J} \, ,
\label{e:effective_Npoints}
\end{align}
where $N_{\rm pts.}$ is the effective number of data points in the weighted $\chi^2$  method. 
In Sec.~\ref{sec:unweighted_chi2}, we provide a comparison of our final fit results in the weighted $\chi^2$ method to the fit results obtained with the usual definition of $\chi^2$ (\ref{e:chi2}). 
(Such a scenario corresponds to simply choosing $w=1$.)

\begin{figure}[t]
\centering
\includegraphics[width=0.85\textwidth]{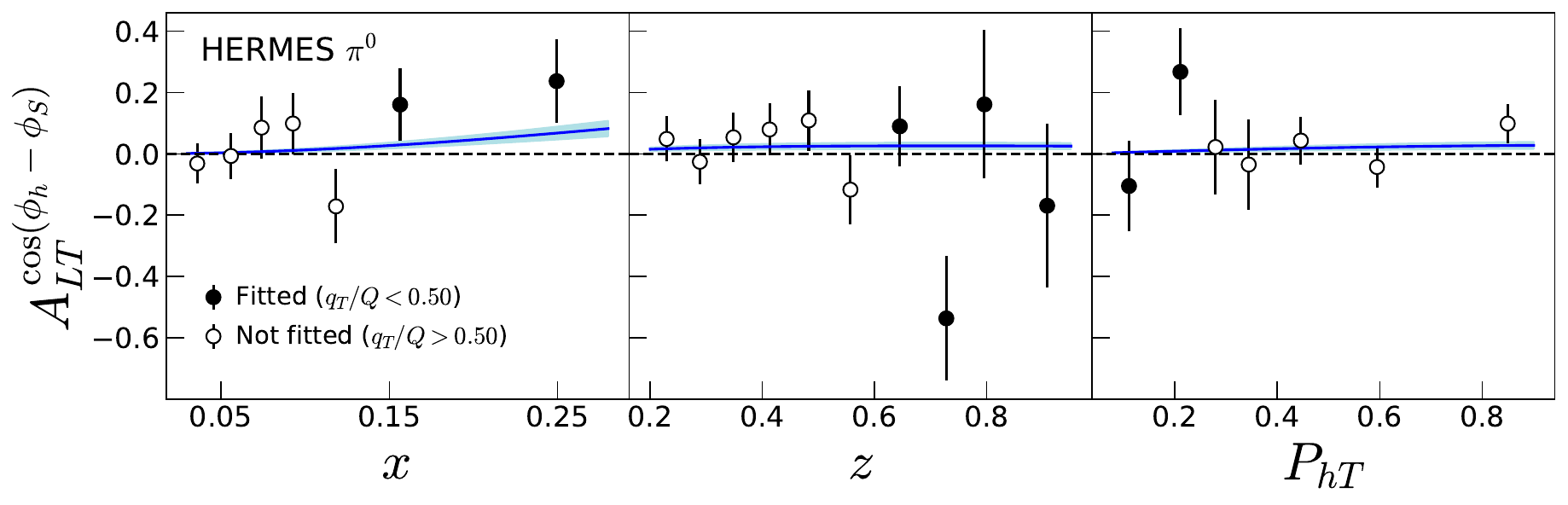}
\caption{HERMES data for $\pi^0$ compared with our theory curves at $1$-$\sigma$ CL.  The dark (open) points are data that were (were not) included in our fit after the $q_T/Q<0.5$ cut.}
\label{f:hermes_pi0}
\end{figure}

\begin{figure}[H]
\centering
\includegraphics[width=0.75\textwidth]{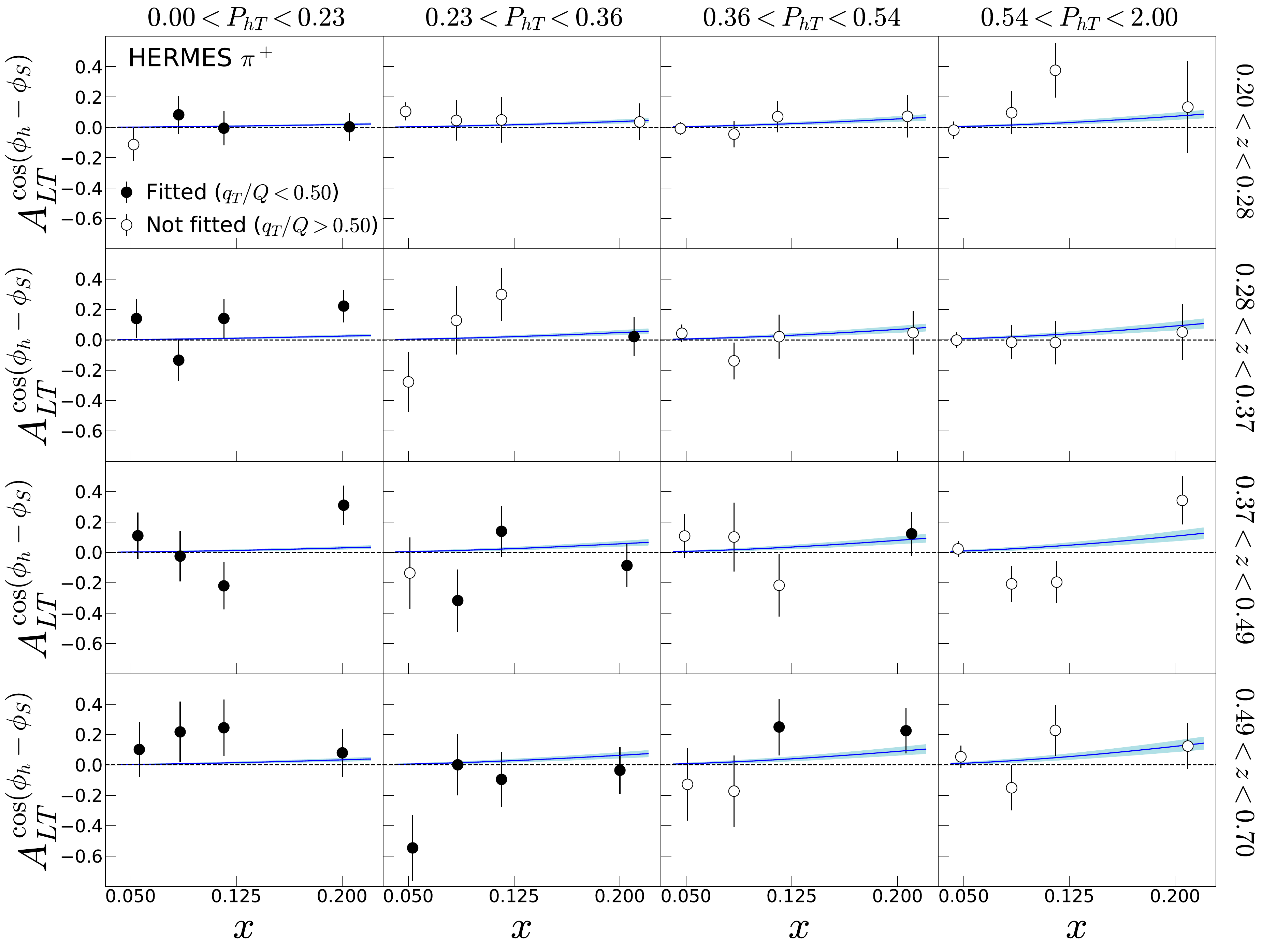}
\caption{HERMES data for $\pi^+$ compared with our theory curves at $1$-$\sigma$ CL.  The dark (open) points are data that were (were not) included in our fit after the $q_T/Q<0.5$ cut.}
\label{f:hermes_pi+}
\end{figure}

\begin{figure}[H]
\centering
\includegraphics[width=0.75\textwidth]{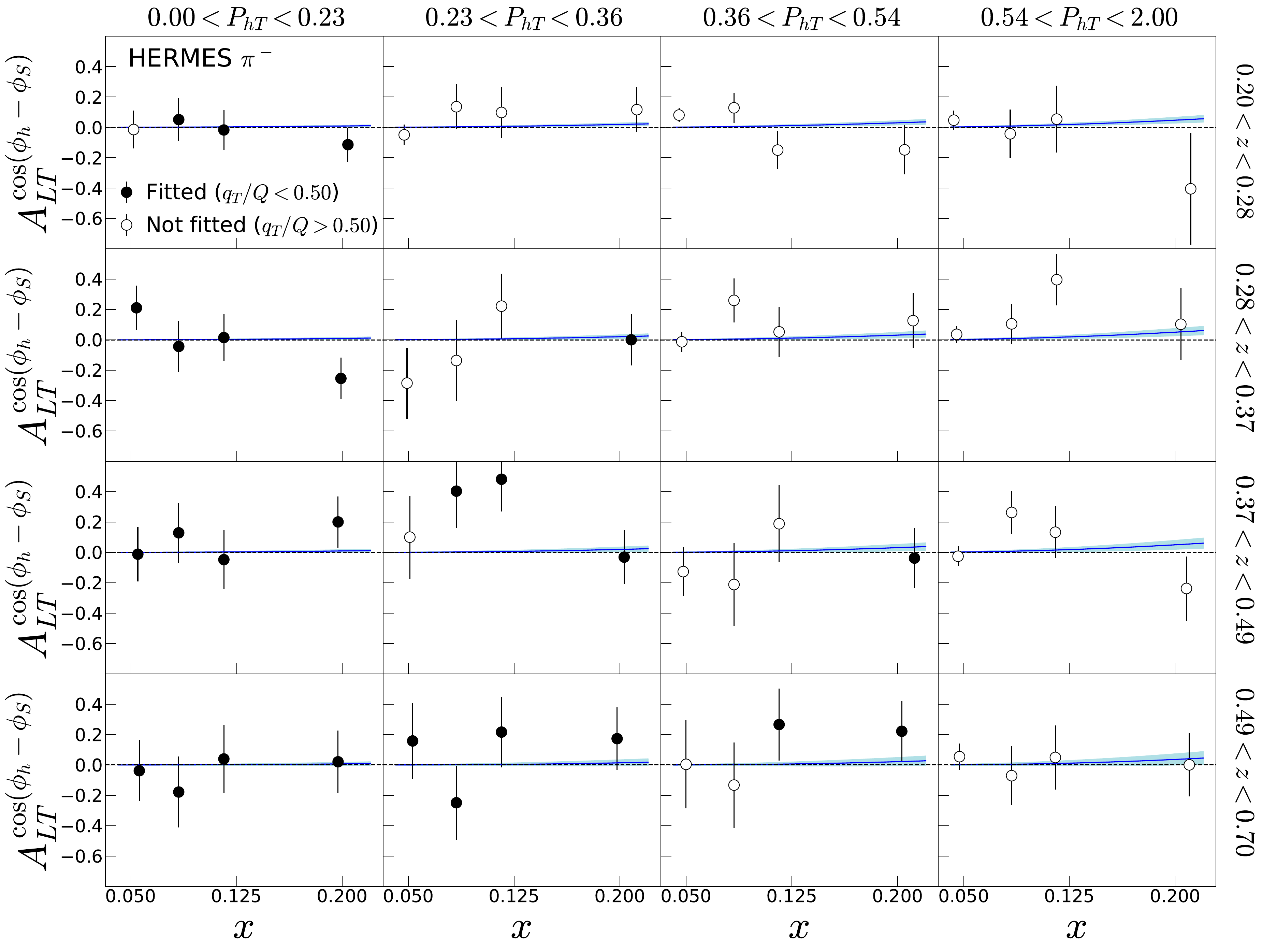}
\caption{HERMES data for $\pi^-$ compared with our theory curves at $1$-$\sigma$ CL.  The dark (open) points are data that were (were not) included in our fit after the $q_T/Q<0.5$ cut.}
\label{f:hermes_pi-}
\end{figure}

\begin{figure}[htbp!]
\centering
\includegraphics[width=0.78\textwidth]{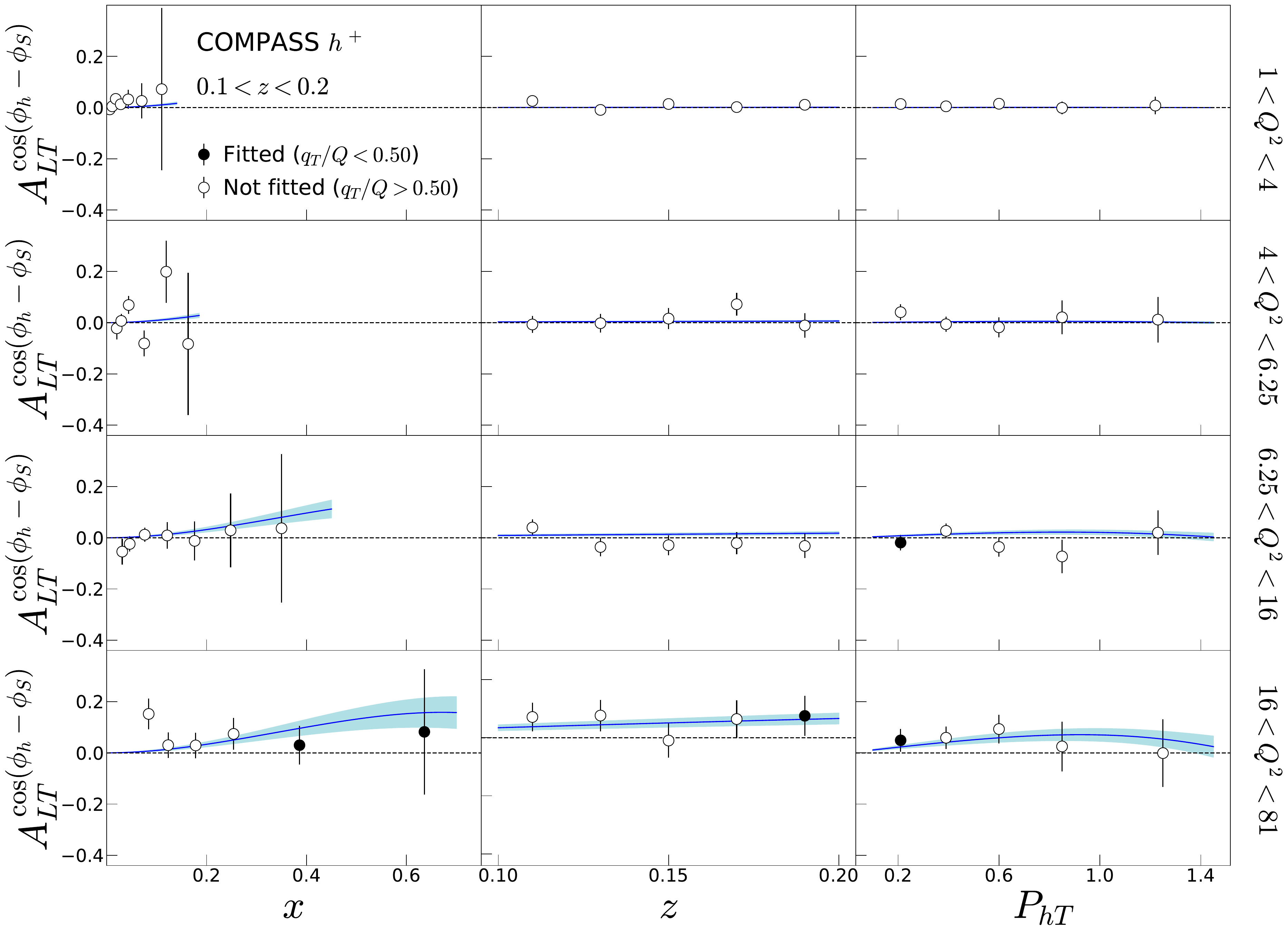}
\caption{COMPASS data for $h^+$ for the bin $0.1 < z < 0.2$ compared with our theory curves at $1$-$\sigma$ CL.  The dark (open) points are data that were (were not) included in our fit after the $q_T/Q<0.5$ cut.}
\label{f:compass_h+_smallbin}
\end{figure}

\begin{figure}[H]
\centering
\includegraphics[width=0.78\textwidth]{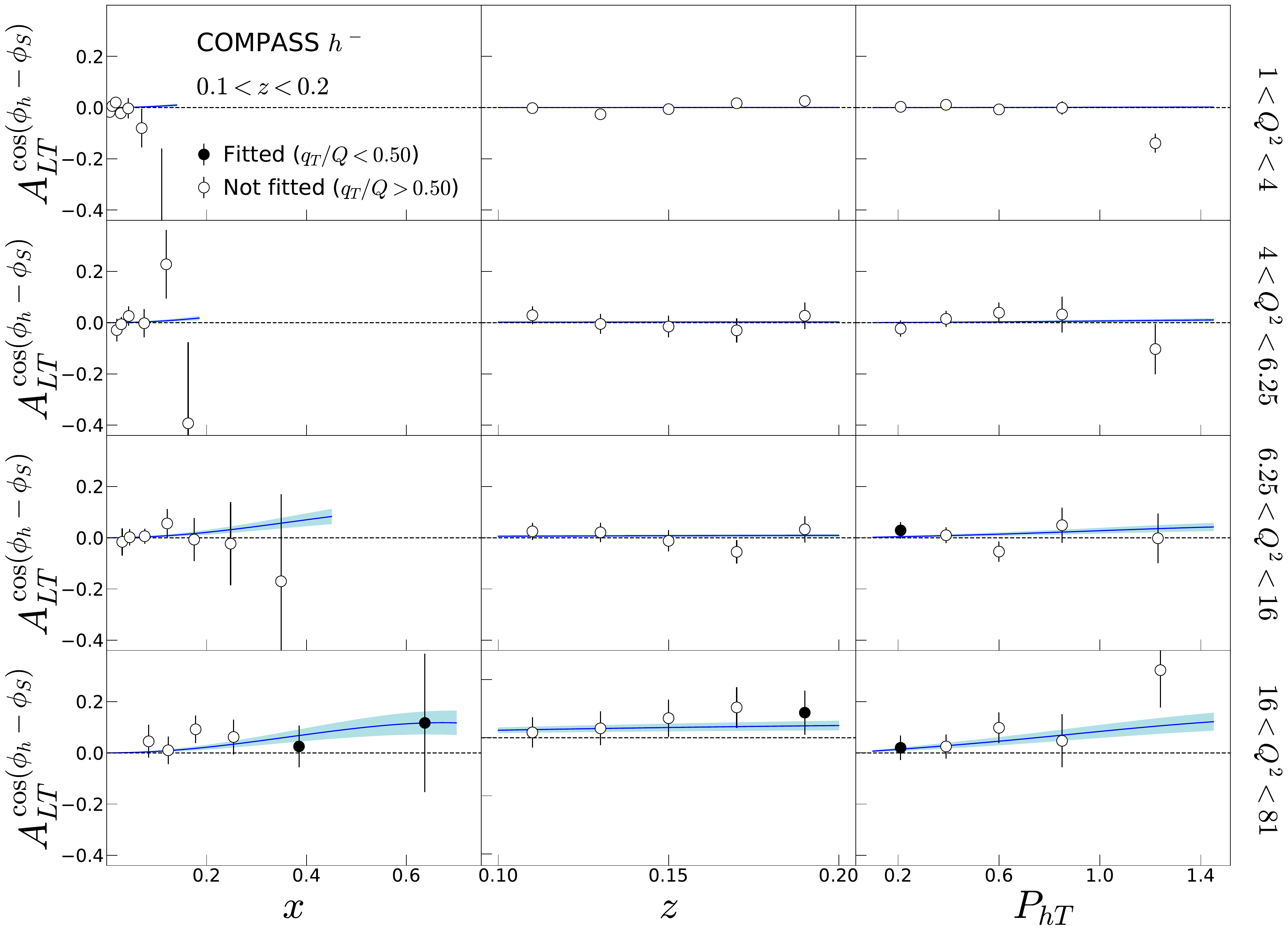}
\caption{COMPASS data for $h^-$ for the bin $0.1 < z < 0.2$ compared with our theory curves at $1$-$\sigma$ CL.  The dark (open) points are data that were (were not) included in our fit after the $q_T/Q<0.5$ cut.}
\label{f:compass_h-_smallbin}
\end{figure}

\begin{figure}[H]
\centering
\includegraphics[width=0.78\textwidth]{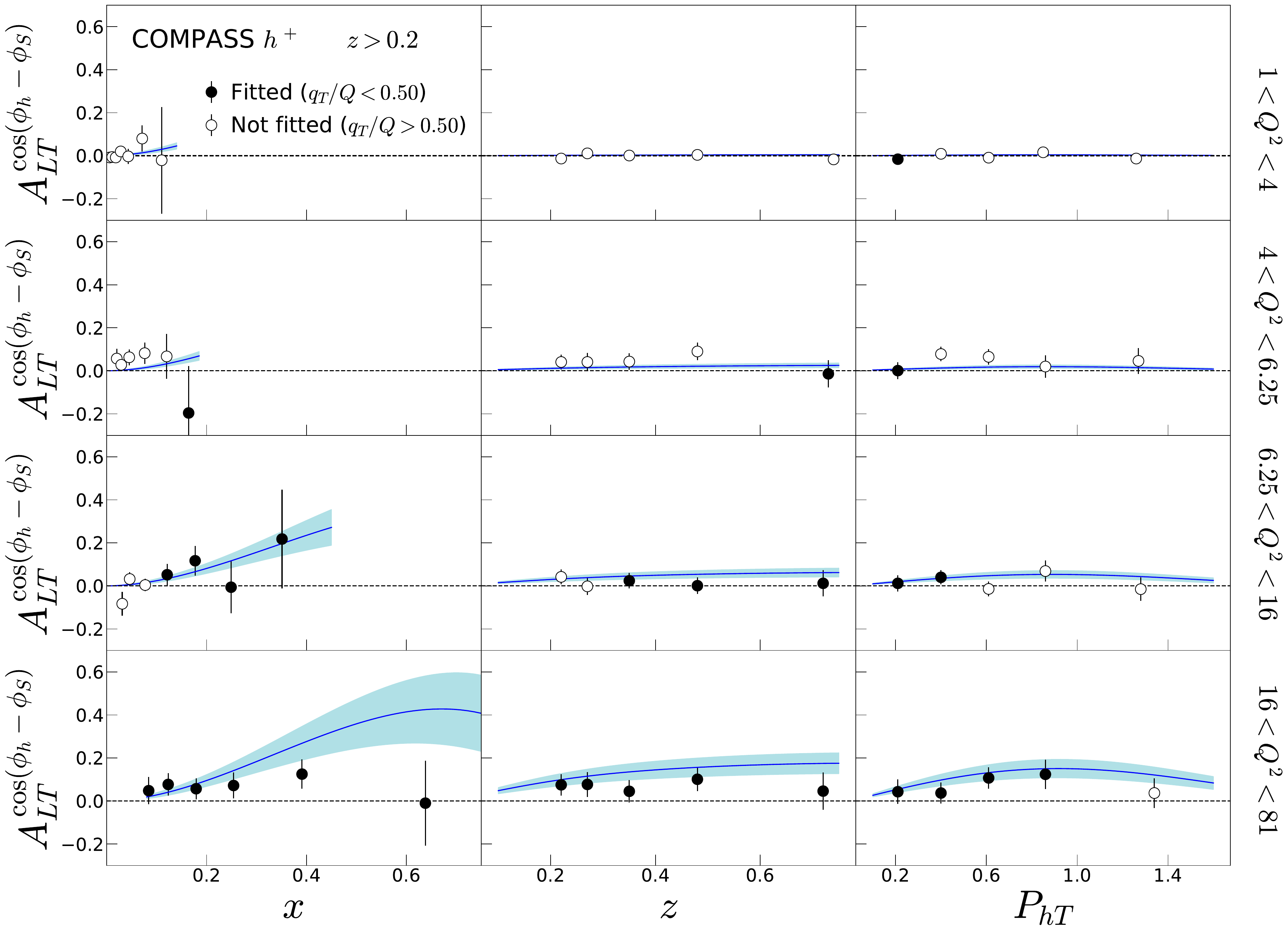}
\caption{COMPASS data for $h^+$ for the bin $z > 0.2$ compared with our theory curves at $1$-$\sigma$ CL.  The dark (open) points are data that were (were not) included in our fit after the $q_T/Q<0.5$ cut.}
\label{f:compass_h+_bigbin}
\end{figure}

\begin{figure}[htbp!]
\centering
\includegraphics[width=0.78\textwidth]{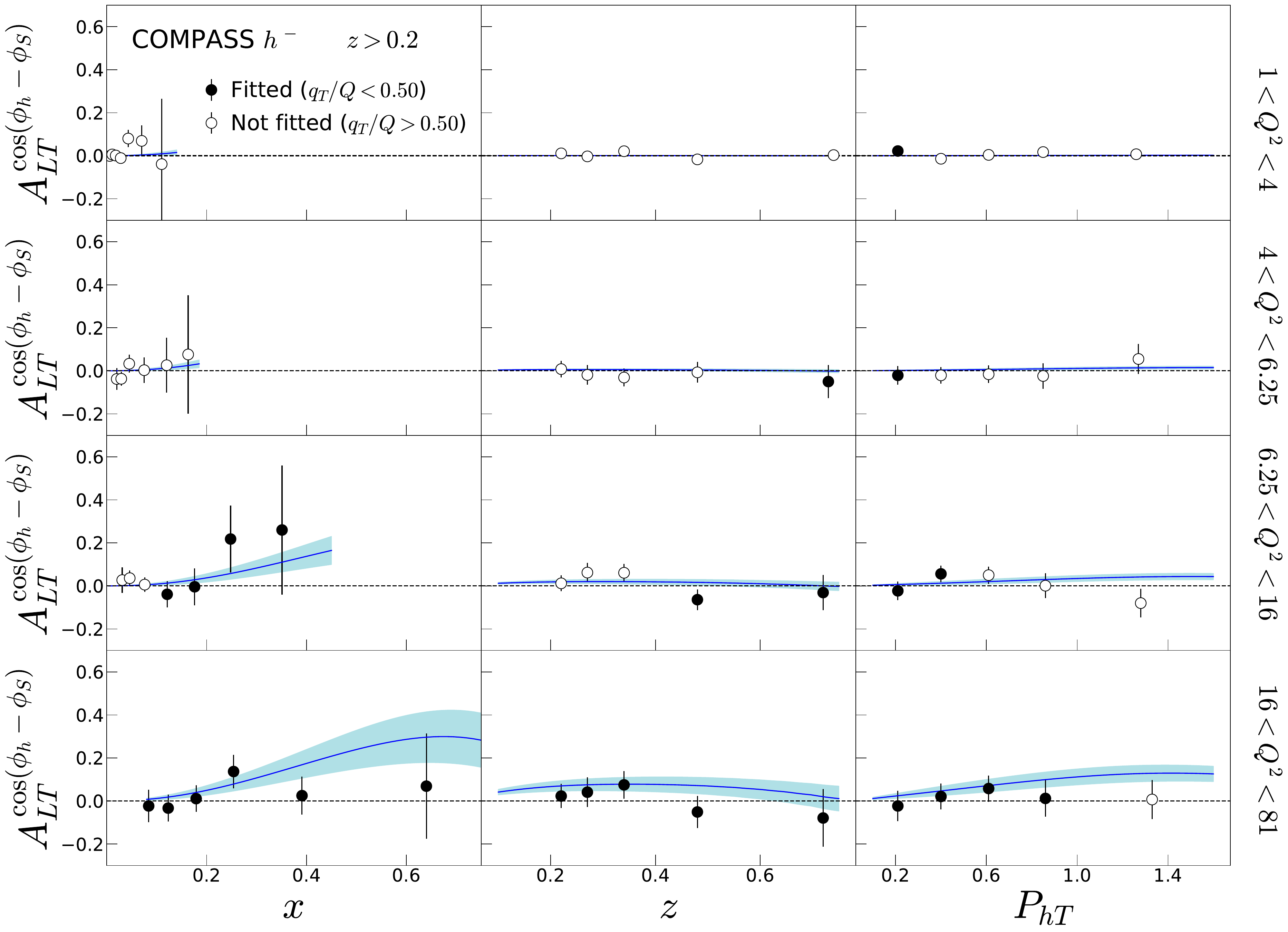}
\caption{COMPASS data for $h^-$ for the bin $z > 0.2$ compared with our theory curves at $1$-$\sigma$ CL.  The dark (open) points are data that were (were not) included in our fit after the $q_T/Q<0.5$ cut.}
\label{f:compass_h-_bigbin}
\end{figure}

\begin{figure}[htbp!]
\centering
\includegraphics[width=0.42\textwidth]{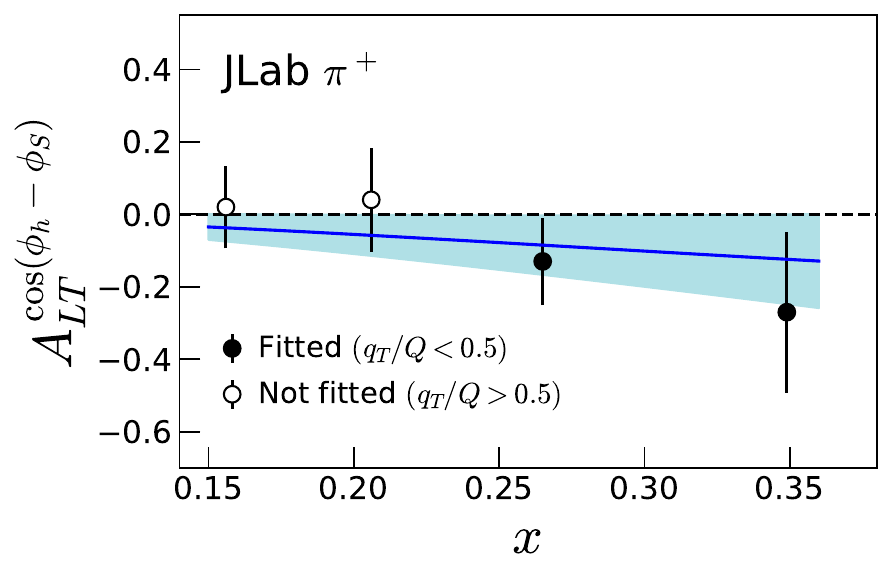}
\includegraphics[width=0.42\textwidth]{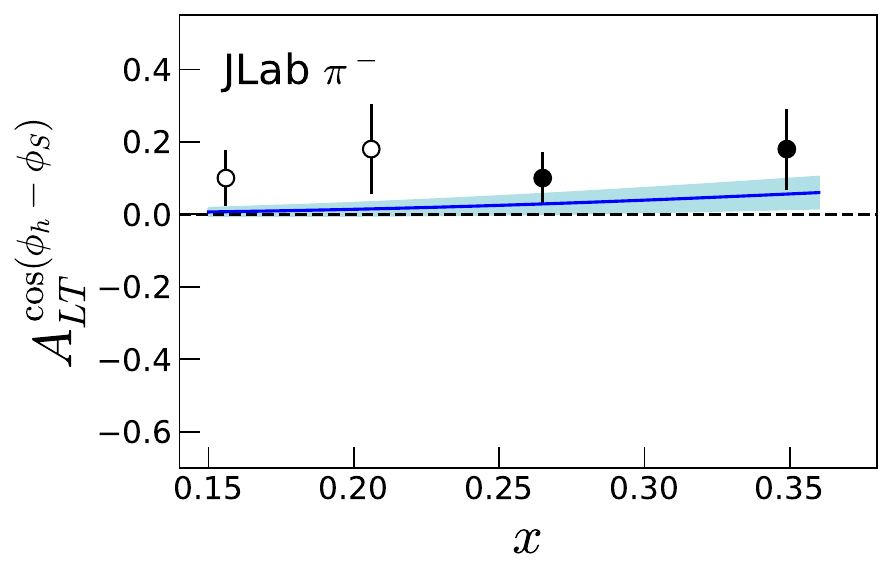}
\caption{JLab data for $\pi^+$ (left) and $\pi^-$ (right) compared with our theory curves at $1$-$\sigma$ CL.  The dark (open) points are data that were (were not) included in our fit after the $q_T/Q<0.5$ cut.}
\label{f:jlab_pi+_pi-}
\end{figure}

In Figs.~\ref{f:hermes_pi0}--\ref{f:jlab_pi+_pi-}
we plot our theoretical curves against the experimental data, with Figs.~\ref{f:hermes_pi0},~\ref{f:hermes_pi+}, and~\ref{f:hermes_pi-} displaying the HERMES data for $\pi^0$, $\pi^+$, and $\pi^-$, Figs.~\ref{f:compass_h+_smallbin}, \ref{f:compass_h-_smallbin} (\ref{f:compass_h+_bigbin}, \ref{f:compass_h-_bigbin}) giving the COMPASS data for the $z$ interval $0.1<z<0.2$ ($z>0.2$) for $h^{+}$ and $h^-$, and Fig.~\ref{f:jlab_pi+_pi-} showing the JLab data for $\pi^+$ and $\pi^-$.  In Table~\ref{table:summary_chi2_w}, we summarize $\chi_w^2/N_{\rm pts.}$ for each data set along with the global $\chi^{2}_w/N_{\rm pts.}$. 
For HERMES, the $\chi^{2}_w/N_{\rm pts.}=1.20$ for $\pi^+$ and $\chi^{2}_w/N_{\rm pts.}=0.88$ for $\pi^-$, thus indicating good agreement between our theory and the data. 
{\flushleft On} the other hand, $\chi^{2}_w/N_{\rm pts.}=1.94$ for $\pi^0$, thus implying that the agreement with our theory is just fair for this data set. The reason for the larger $\chi^{2}_w/N_{\rm pts.}$ most likely is the few points that deviate from the overall trend of the data. For COMPASS, $\chi^{2}_w/N_{\rm pts.}=0.97$ for $h^+$  and $\chi^{2}_w/N_{\rm pts.}=0.71$ for $h^-$ data. For JLab, $\chi^{2}_w/N_{\rm pts.}=0.31$ for  $\pi^+$ data and $\chi^{2}_w/N_{\rm pts.}=1.13$ for  $\pi^-$. These values suggest strong compatibility between our theory and the data. We comment that the theoretical uncertainty bands in certain kinematic bins, especially for HERMES, appear to be rather small.  To explore this observation, we performed a fit exclusively of the HERMES and JLab data and found several of the COMPASS data points in the highest $Q^2$ bin do not overlap with the  error bands of the theory ``prediction'' from such a fit.  That is, the typical $g_{1T}$ functions that describe HERMES (and JLab) only are too large in certain kinematic regions to describe COMPASS. Consequently, when we perform our global fit, the theory calculation for HERMES falls within a very limited range.

\begin{figure}[H]
\centering
\includegraphics[width=0.9\textwidth]{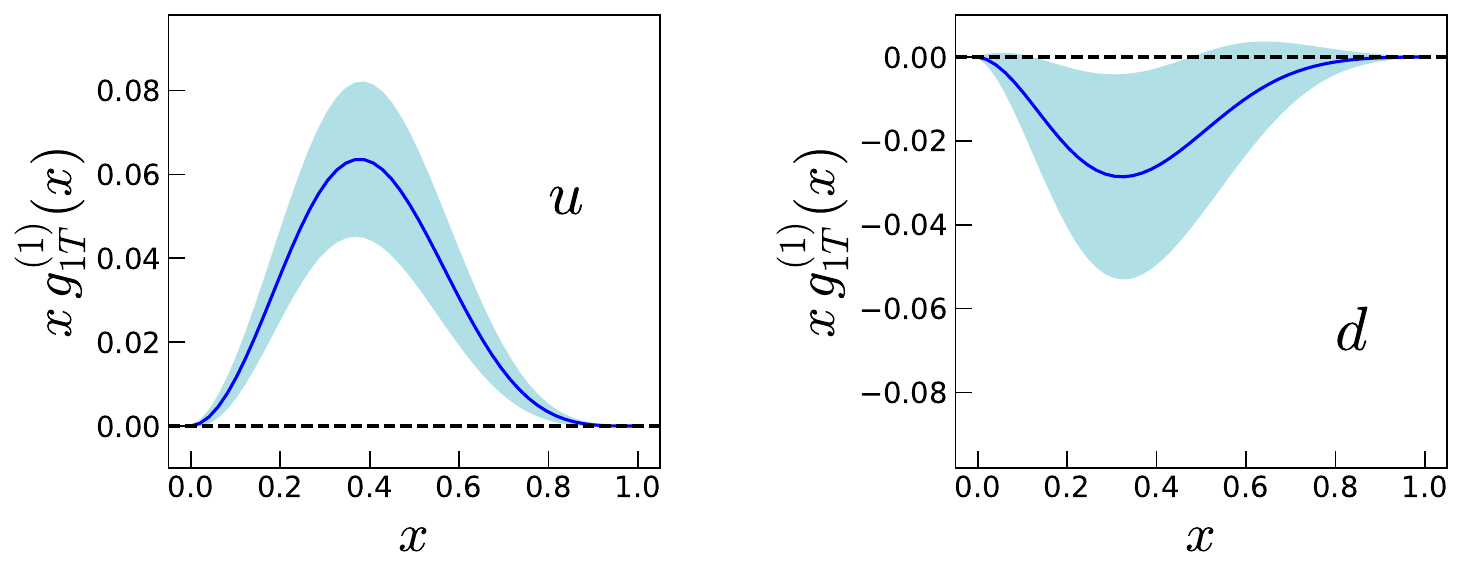}
\caption{Main global fit results for $xg_{1T}^{(1)}(x)$ at $Q^{2}$ = 4 ${\rm GeV}^2$ for up quarks (left) and down quarks (right) obtained in the weighted $\chi^2$ method.}
\label{f:main_fit}
\end{figure}

In Fig.~\ref{f:main_fit} we show our final results for $x \, g^{(1)}_{1T}(x)$ for the up quarks (left panel) and the down quarks (right panel) at $Q^{2}=4 \,\rm GeV^2$. (In Appendix~\ref{app_a} we show a plot with all the replicas.) We emphasize that this is the very first information on $g_{1T}(x,\vec{k}_\perp^2)$ from experimental data, which we have obtained from an analysis of the world SIDIS data on $A_{LT}^{\cos(\phi_h-\phi_S)}$.
Figure~\ref{f:main_fit} indicates that $g_{1T}$ for the up quark is positive and for the down quark is negative, although with large error bands for the latter. 
This is most likely because, even though JLab does have neutron data, the $\pi^+$ errors are larger compared to those for $\pi^-$ (see Fig.~\ref{f:jlab_pi+_pi-}), so one cannot achieve as precise an extraction for the down quark $g_{1T}$.  Additional data from JLab on a neutron target, or COMPASS on a deuteron target, would be needed to obtain a better flavor separation.
Nonetheless, the first prominent qualitative feature that we observe here is that our results are compatible (to some extent) with the large-$N_{c}$ approximation (\ref{e:largeNc_g1T}), which implies that $g_{1T}^u$ and $g_{1T}^d$ have relative signs. Recall that such was the conclusion also from lattice QCD as well as calculations in constituent quark models. Note that we mention ``to some extent" because of the relative sizes of the two distributions. 
The values of the parameters from our fit are $N^{u} = 0.026 \pm 0.007$, $N^{d} =-0.012 \pm 0.010$, and $\alpha =1.9 \pm 0.6$, so the magnitude of $g_{1T}^u$ does slightly overlap with that for $g_{1T}^d$.  We will present results for a fit that imposes the large-$N_c$ constraint $g_{1T}^u = -g_{1T}^d$ in the next subsection.

\subsection{Large-$N_c$ and WW-type approximation results}
\begin{figure}[t]
\centering
\includegraphics[width=0.9\textwidth]{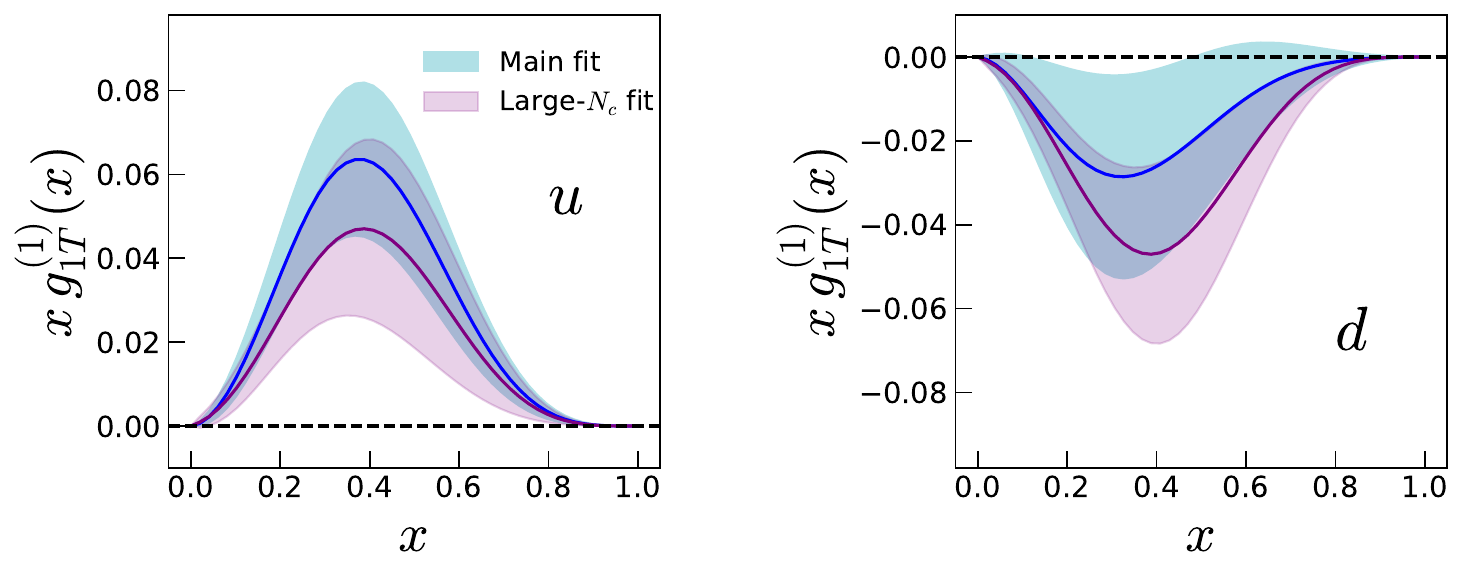}
\caption{Comparison of our main extraction of $xg_{1T}^{(1)}(x)$ at $Q^{2}$ = 4 ${\rm GeV}^2$ for up quarks (left) and down quarks (right) with the results obtained by imposing the large-$N_c$ approximation on the fit.  Both functions were obtained using the weighted $\chi^2$ method.}
\label{f:main_fit_vs_LargeNc}
\end{figure}
In this subsection, we study how well the experimental data support the large-$N_c$ and WW-type approximations for $g_{1T}$ (see Sec.~\ref{sec:theory}) by separately imposing these constraints and analyzing the $\chi^{2}_w/N_{\rm pts.}$ for each case.

To perform the fit in the large-$N_c$ approximation, we set $g^{(1)d}_{1T}(x) = - g^{(1)u}_{1T}(x)$. Therefore, this analysis involves fitting two free parameters, $N^u$ and $\alpha$, and we also carry this out using the weighted $\chi^2$ method.
In Fig.~\ref{f:main_fit_vs_LargeNc}, we provide a comparison of our final fit from the previous subsection with the large-$N_c$ results. We find that the large-$N_c$ curves and our final results overlap rather well within error bands. In Table~\ref{table:summary_chi2_w}, we also compare the $\chi^{2}_w/N_{\rm pts.}$ for the two cases. We observe that our main fit gives a better global $\chi^{2}_w/N_{\rm pts.}$ than the large-$N_c$ fit. 
However, one must keep in mind that these $\chi^{2}_w/N_{\rm pts.}$ values correspond to the central curves of the theory. Since every replica has its own $\chi^{2}_w/N_{\rm pts.}$ value, it is actually more informative to make a comparison between the $\chi^{2}_w/N_{\rm pts.}$-distribution for our main fit and the large-$N_c$ fit to determine which of the two approaches produces better results (if at all). In Fig.~\ref{f:chi2_distribution}, we show histograms of $\chi^{2}_w/N_{\rm pts.}$ for each data set for the two cases. We observe a significant overlap of the $\chi^{2}_w/N_{\rm pts.}$-distributions, especially for (but not limited to) the COMPASS data sets for $h^\pm$ and for the JLab data set for $\pi^+$. This implies that, although there is a slight preference for $g_{1T}$ to violate the large-$N_c$ approximation, there is actually no statistically significant difference between the two fits (main and large-$N_c$).  That is, the large-$N_c$ approximation is consistent with the experimental data.

\begin{figure}[t]
\centering
\includegraphics[width=0.9\textwidth]{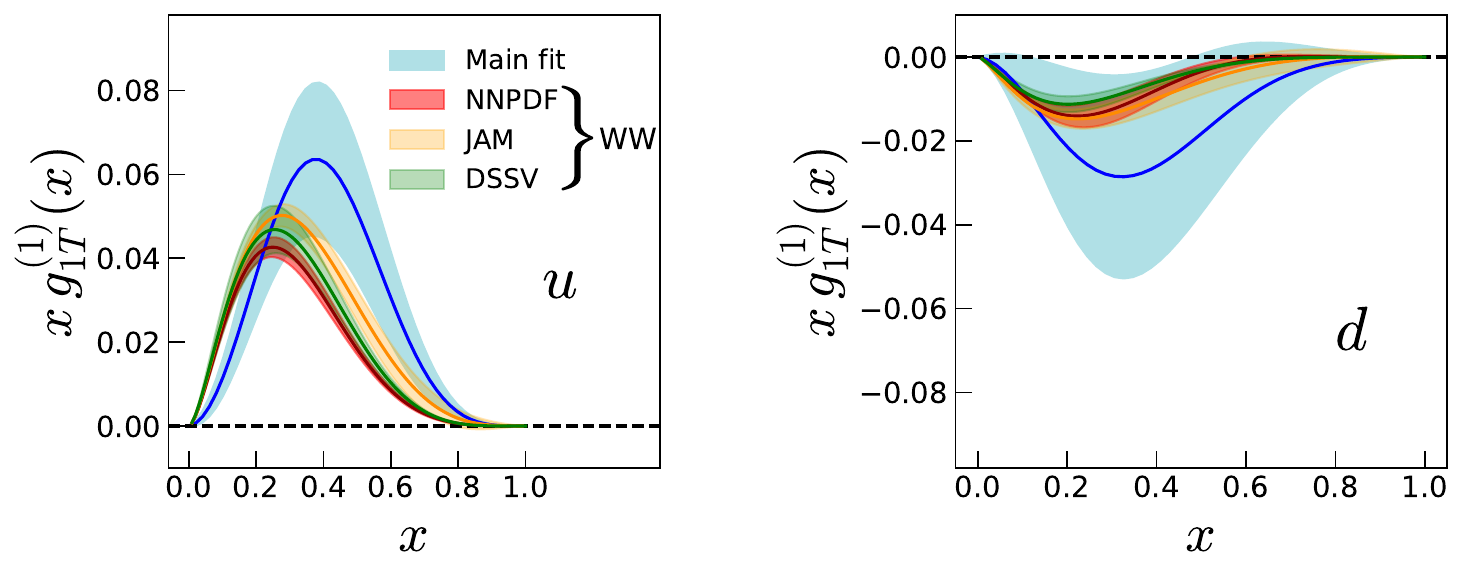}
\caption{Comparison of our main extraction of $xg_{1T}^{(1)}(x)$ at $Q^{2}$ = 4 ${\rm GeV}^2$ for up quarks (left) and down quarks (right) with the results obtained from the calculation of $xg^{\rm WW}_{1T}(x)$ using Eq.~(\ref{e:WWtype}) with $g_1(x)$ taken from NNPDF~\cite{Nocera:2014gqa}, JAM~\cite{Ethier:2017zbq}, and DSSV~\cite{DeFlorian:2019xxt}.}
\label{f:main_fit_vs_WW}
\end{figure}

\begin{figure}[H]
\centering
\includegraphics[width=0.9\textwidth]{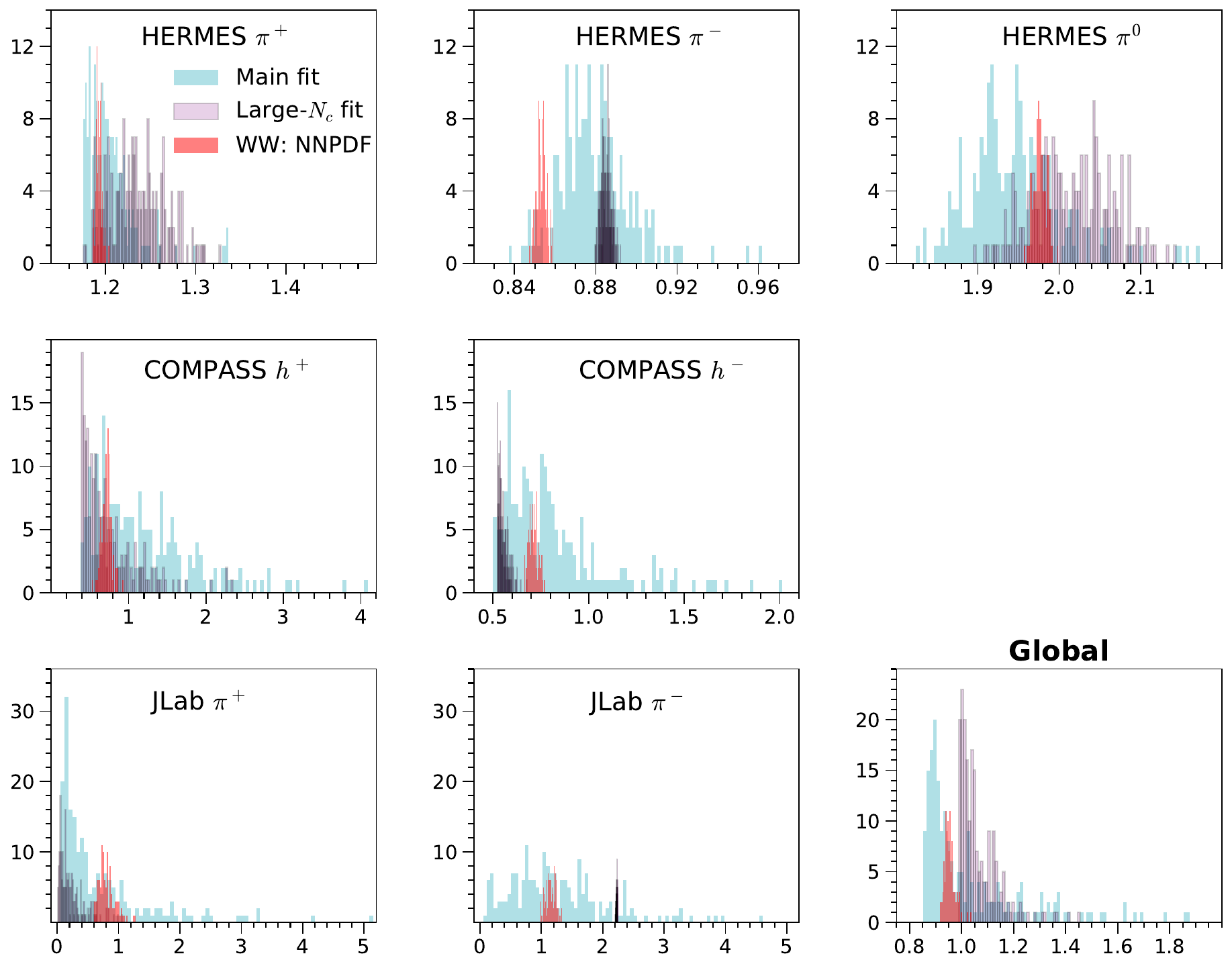}
\caption{Distribution of $\chi^{2}_w/N_{\rm pts.}$ for our main fit results with those obtained in the large-$N_c$ and the WW-type approximations. For the latter, we have shown as the representative case results using NNPDF. The results do not qualitatively change with using JAM or DSSV.}
\label{f:chi2_distribution}
\end{figure}

In Fig.~\ref{f:main_fit_vs_WW}, we provide a comparison of our main fit with a calculation of $g_{1T}^{(1)}(x)$ using the WW-type approximation~(\ref{e:WWtype}) with $g_1(x)$ taken from NNPDF~\cite{Nocera:2014gqa}, JAM~\cite{Ethier:2017zbq}, and DSSV~\cite{DeFlorian:2019xxt}. Overall, we observe a qualitative agreement with the WW-type approximation in terms of the general behavior of $g_{1T}$. However, there are slight differences that may hint at a violation of the WW-type approximation for the up quark.
As for the down quark, there seems to be an agreement between our main fit and the WW-type approximation, admittedly because of the presence of large uncertainties in our extraction. More precise data will certainly be needed to affirm the degree of violation (if any) of the WW-type approximation.  This is especially important to resolve since a clear breaking of the WW-type approximation would be a direct signal of quark-gluon-quark correlations in the nucleon~\cite{Avakian:2007mv,Accardi:2009au,Kanazawa:2015ajw,Scimemi:2018mmi}. In Table~\ref{table:summary_chi2_w}, we compare the $\chi^{2}_w/N_{\rm pts.}$ for our main fit with that from calculating $A_{LT}^{\cos(\phi_h-\phi_S)}$ in the WW-type approximation. (We note that the WW-type approximation results are {\it not} from a fit since $g_{1T}^{(1)}(x)$ is fixed by using Eq.~(\ref{e:WWtype}) with $g_1(x)$ taken from NNPDF, JAM, or DSSV, and the width $\langle k_\perp^2\rangle |_{g_{1T}}$ is fixed to be the same as the one used in our main fit.)
We not only obtain the same or somewhat better $\chi^2$ for each of the data sets as the WW-type approximation, but the global $\chi^{2}_w/N_{\rm pts.}$ for our main fit is better than that for the WW-type approximation.
However, as before, this does not imply that our fit is favored over the WW-type approximation. As shown in Fig.~\ref{f:chi2_distribution}, the statistical spread of the $\chi^{2}_w/N_{\rm pts.}$-distribution for our fit results are rather large and hence overlaps significantly with the $\chi^{2}_w/N_{\rm pts.}$-distribution from the WW-type approximation. This confirms that at present the WW-type approximation is compatible with the experimental data.

\begin{table}[b]
\setlength{\tabcolsep}{4pt}
\renewcommand{\arraystretch}{1.6}
\centering
\begin{tabular}{| c | c | c | c | c | c | c |}
\hline
\multicolumn{6}{|c|}{{\bf Summary of} $\boldsymbol{\chi^{2}_w/N_{\rm pts.}}$} \\
\hline
Data set \, &  $\chi^{2}_w/N_{\rm pts.}\big |_{\rm Main}$ \, & $\chi^{2}_w/N_{\rm pts.}\big |_{{\rm Large}{\text -}N_{c} }$ & \, $\chi^{2}_w/N_{\rm pts.}\big |_{\rm NNPDF}$ & \, $\chi^{2}_w/N_{\rm pts.}\big |_{\rm JAM}$ & \, $\chi^{2}_w/N_{\rm pts.}\big |_{\rm DSSV}$ \\
\hline 
\hline 
$\rm{HERMES}\; \pi^{+}$   &  1.20 & 1.23  &  1.19  &  1.19 &  1.19\\
\cline{1-6}
$\rm{HERMES}\; \pi^{-}$   &  0.88 & 0.88  &  0.85  &  0.85 &  0.85\\
\cline{1-6}
$\rm{HERMES}\; \pi^{0}$   &  1.94 & 2.01  &  1.98  &  1.95 &  1.96\\
\cline{1-6}
${\rm COMPASS}\; h^{+}$     &  0.97 & 0.51  &  0.71  &  1.02 &  0.89\\
\cline{1-6}
${\rm COMPASS}\; h^{-}$     &  0.71 & 0.53  &  0.71 &  0.81 &  0.80\\
\cline{1-6}
$\rm{JLab}\; \pi^{+}$   &  0.31 & 0.06  &  0.81 &  0.78  &  0.96\\
\cline{1-6}
$\rm{JLab}\; \pi^{-}$   &  1.13 & 2.23  &  1.15 &  0.93  &  0.93\\
\cline{1-6}
{\bf Global}                & {\bf 0.86}  & {\bf 0.99}  &  {\bf 0.95} &  {\bf 0.94}  &  {\bf 0.97}\\ 
\hline
\end{tabular}
\caption{Summary of the $\chi^{2}_w/N_{\rm pts.}$ for our main fit and for the large-$N_{c}$ and the WW-type approximation results in the weighted $\chi^{2}$ method.  For the latter, we separately give results using $g_1(x)$ from NNPDF, JAM, and DSSV.}
\label{table:summary_chi2_w}
\end{table}

\subsection{Comparison with lattice QCD}
\begin{figure}[t]
\centering
\includegraphics[width=0.5\textwidth]{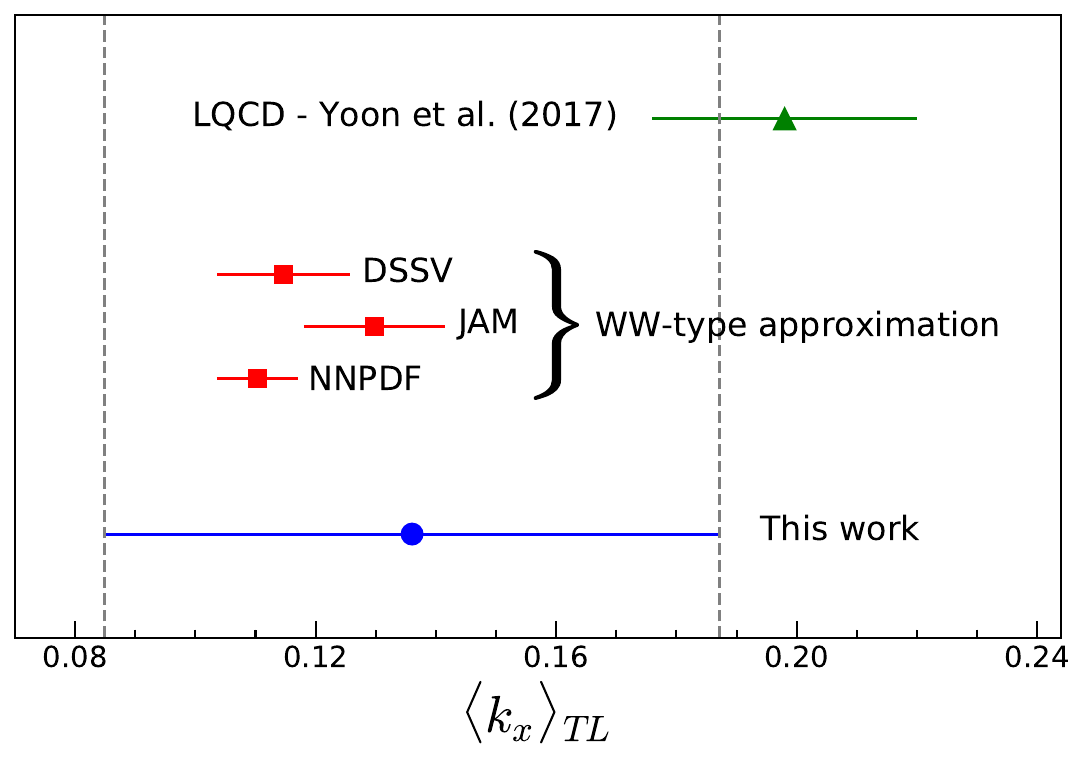}
\caption{Comparison of the worm-gear shift $\langle k_{x} \rangle _{TL}$ calculated using our main fit (blue circle) with those obtained in the WW-type approximation (red squares) as well as lattice QCD (green triangle). The latter is the rightmost data point from Fig.~13 of Ref.~\cite{Yoon:2017qzo} in the domain-wall fermion (DWF) scheme. The phenomenological results are at $Q^{2}$ = 4 ${\rm GeV}^2$.}
\label{f:kx}
\end{figure}

In this section, we explore the so-called worm-gear shift $\langle k_{x} \rangle _{TL}$ for $g_{1T}$, 
\begin{align}
\big[\langle k_{x} \rangle _{TL}\big] \!(Q^{2}) & \equiv M \, \dfrac{\int^{1}_0 dx \, \Big [ g^{(1)u}_{1T} (x, Q^{2}) - g^{(1)d}_{1T} (x, Q^2) \Big ]}{\int^{1}_0  dx \, \Big [f^{u}_{1} (x, Q^{2}) - f^{d}_{1} (x, Q^2) \Big ]} \, .
\label{e:kx}
\end{align}
The motivation for calculating this quantity is that it has also been computed using lattice QCD~\cite{Hagler:2009mb,Musch:2010ka,Yoon:2017qzo}. The results are shown in Fig.~\ref{f:kx} for our main fit, the WW-type approximation, and a lattice point from Ref.~\cite{Yoon:2017qzo} (namely, the rightmost data point from Fig.~13 of Ref.~\cite{Yoon:2017qzo} in the domain-wall fermion (DWF) scheme).  We find consistency between lattice and our main fit as well as the WW-type approximation and our main fit, but a slight discrepancy between lattice and the WW-type approximation, again perhaps hinting at a breaking of the latter. A similar value for the worm-gear shift in the WW-type approximation was found in Ref.~\cite{Scimemi:2018mmi}.  We also note that from the large-$N_c$ fit (Fig.~\ref{f:main_fit_vs_LargeNc}), $\langle k_{x} \rangle _{TL} = 0.14 \pm 0.05$, similar to that of our main fit.  We mention some caveats in making a direct comparison between phenomenology and lattice for the worm-gear shift. The scale for the main fit and WW-type approximation points is $Q^2 = 4\,{\rm GeV^2}$.  For lattice, there is not a definitive scale, but it can be approximated by the inverse lattice spacing, from which one finds $Q^2\approx 5.52\,{\rm GeV^2}$ from the spacing $a=0.084\,{\rm fm}$ used in Ref.~\cite{Yoon:2017qzo} for the DWF scheme.  In addition, the full correspondence between the theoretical and lattice calculations of $\langle k_{x} \rangle _{TL}$ occurs in the $\hat{\zeta}\to \infty, b_T\to 0$ limit of the lattice result, where $\hat{\zeta}$ is a Collins-Soper type parameter and $b_T$ is the separation between the quark fields~\cite{Yoon:2017qzo}.  The specific lattice point we compare with in our Fig.~\ref{f:kx} has $\hat{\zeta}\approx 0.42$ and $b_T=0.34\,{\rm fm}$, which were the closest values used in Ref.~\cite{Yoon:2017qzo} to the aforementioned limit.  Nevertheless, the dependence of $\langle k_{x} \rangle _{TL}$ on these lattice parameters seems very mild~\cite{Yoon:2017qzo}.  It is encouraging at this stage that experimental data and lattice are in reasonable agreement.

\subsection{Results in the unweighted $\chi^2$ method}
\label{sec:unweighted_chi2}

\begin{figure}[t]
\centering
\includegraphics[width=0.9\textwidth]{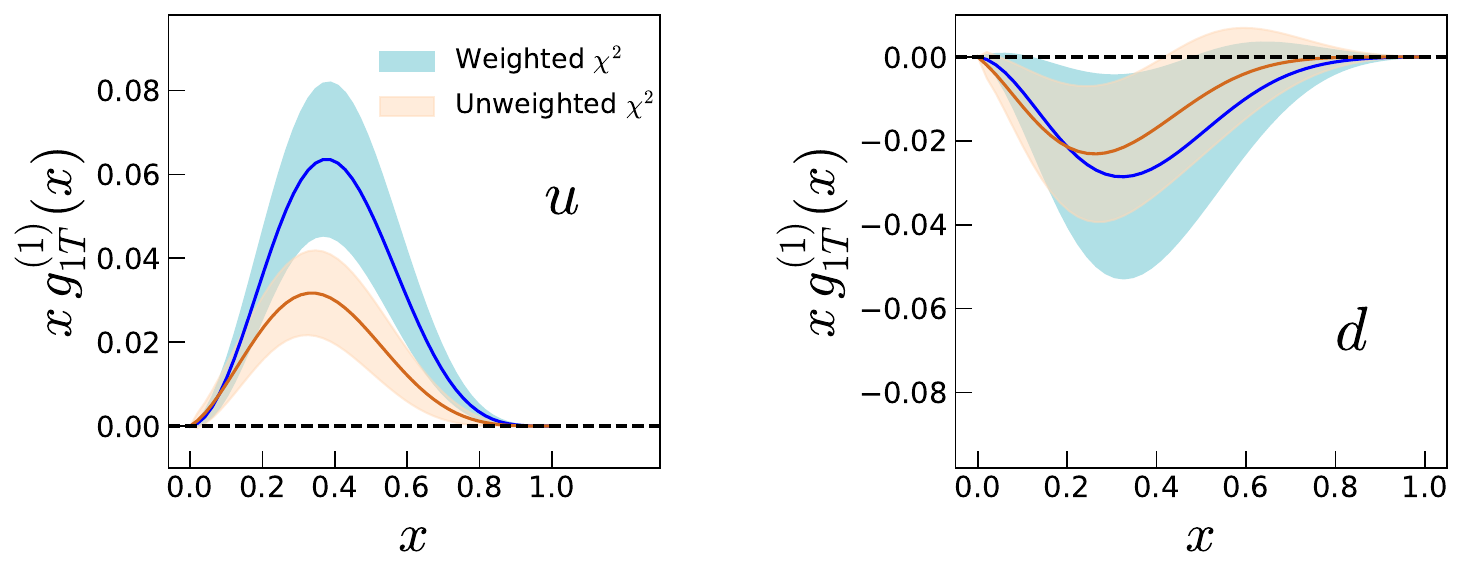}
\caption{Global fit results for $xg_{1T}^{(1)}(x)$ at $Q^{2}$ = 4 ${\rm GeV}^2$ for up quarks (left) and down quarks (right) obtained in the unweighted $\chi^2$ method (light brown) compared with our main fit (light blue).}
\label{f:weighted_vs_unweighted}
\end{figure}

In this section, we discuss what happens to $g_{1T}$ if we do not weight the JLab data at all, which corresponds to $w=1$ in Eq.~(\ref{e:chi2_w}).
In Fig.~\ref{f:weighted_vs_unweighted}, we compare our main fit results to the ones with $w=1$.
We see that the impact is significantly larger on the up quark, while for the down quark the two cases overlap rather well within error bands.
We also notice that the JLab data for $\pi^-$ is not described particularly well if we do not weight the $\chi^2$ (cf.~the relevant entry on the second column of Table~\ref{table:summary_chi2} with that of Table~\ref{table:summary_chi2_w}). 
Since JLab data are for neutrons, the $\pi^-$ channel, which has somewhat precise data, has the most impact on the {\it proton} $g_{1T}$ for an {\it up quark} (which is the function plotted in the left panel of Fig.~\ref{f:weighted_vs_unweighted}). We therefore conclude that what we currently have in Fig.~\ref{f:main_fit} is the best fit that one can provide at this stage, certainly because of the strong compatibility that we obtain simultaneously with \textit{all} the data sets. (In Appendix~\ref{app_b} we compare the fit results obtained in the unweighted $\chi^{2}$ method with the results obtained in the (unweighted) large-$N_c$ fit and the WW-type approximation.)

\begin{table}[t]
\setlength{\tabcolsep}{4pt}
\renewcommand{\arraystretch}{1.6}
\centering
\begin{tabular}{| c | c | c | c | c | c | c |}
\hline
\multicolumn{6}{|c|}{{\bf Summary of} $\boldsymbol{\chi^{2}/N_{\rm pts.}}$} \\
\hline
Data set \, &  $\chi^{2}/N_{\rm pts.}\big |_{\rm Main}$ \, &  $\chi^{2}_/N_{\rm pts.}\big |_{{\rm Large}{\text-}N_{c} }$  & \, $\chi^{2}/N_{\rm pts.}\big |_{\rm NNPDF}$ & \, $\chi^{2}/N_{\rm pts.}\big |_{\rm JAM}$ & \, $\chi^{2}/N_{\rm pts.}\big |_{\rm DSSV}$ \\
\hline 
\hline 
$\rm{HERMES}\; \pi^{+}$   & 1.24 &  1.23   &  1.19 &  1.19 &  1.19\\
\cline{1-6}
$\rm{HERMES}\; \pi^{-}$   & 0.89 &  0.88   &  0.85 &  0.85 &  0.85\\
\cline{1-6}
$\rm{HERMES}\; \pi^{0}$   & 2.03 &  2.03   &  1.98 &  1.95 &  1.96\\
\cline{1-6}
${\rm COMPASS}\; h^{+}$     & 0.39 &  0.40   &  0.71 &  1.02 &  0.89\\
\cline{1-6}
${\rm COMPASS}\; h^{-}$      & 0.54 &  0.53  &  0.71 &  0.81 &  0.80\\
\cline{1-6}
$\rm{JLab}\; \pi^{+}$   & 0.42 &  0.15   &  0.81 &  0.78 &  0.96\\
\cline{1-6}
$\rm{JLab}\; \pi^{-}$   & 1.88 &  2.23   &  1.15 &  0.93 &  0.93\\
\cline{1-6}
{\bf Global}                & {\bf 0.83} &  {\bf 0.83}   & {\bf 0.93} &   {\bf 1.02}  & {\bf 0.99}\\ 
\hline
\end{tabular}
\caption{Same as Table~\ref{table:summary_chi2_w} but for the unweighted $\chi^{2}$ method.}
\label{table:summary_chi2}
\end{table}

\section{Summary and outlook}
\label{sec:summary}
In this work, we have performed the first global extraction of the TMD $g_{1T}(x, \vec{k}^{2}_\perp)$ using all experimental measurements available, namely data from HERMES, COMPASS, and JLab on the SIDIS asymmetry $A_{LT}^{\cos(\phi_h-\phi_S)}$. We have used a so-called weighted $\chi^{2}$ method in order to allow the JLab data, which has few data points, to also contribute on the same footing as the HERMES and COMPASS data sets. 
Our main fit indicates that the up quark $g_{1T}$ is positive and the down quark is negative, with the up quark $g_{1T}$ being somewhat larger in magnitude than the down quark. Such a feature is qualitatively compatible with the large-$N_{c}$ approximation and several model calculations.
Actually, the current experimental data cannot rule out the strict large-$N_c$ approximation, namely that $g_{1T}^u = -g_{1T}^d$, as we confirmed by conducting a fit imposing this constraint.
Furthermore, we have provided for the first time a quantitative comparison of $g_{1T}$ from the experimental data with the WW-type approximation and a lattice QCD calculation. Our final fit yields a value for the so-called worm-gear shift that is compatible with lattice QCD.  The agreement between experiment and lattice is encouraging and motivates continued comparisons in the future.
Moreover, our results give a slight indication of a breaking of the WW-type approximation, even though the approximation is still compatible with experimental data. More precise data will be needed to reliably determine how much (if at all) $g_{1T}$ violates the WW-type approximation.  Since such a breaking is directly connected to quark-gluon-quark correlations in the nucleon, it is of great interest to rigorously address this question.  This includes not only further theoretical improvements but also additional measurements of $A_{LT}$ from JLab and COMPASS in SIDIS as well as complimentary measurements at RHIC of vector boson production, which will allow one to study evolution effects in more detail. In the future, we plan to explore the impact of TMD evolution on the present results.

\begin{acknowledgements}
We thank B.~Parsamyan and G.~Schnell for providing us with the COMPASS and HERMES data that was used in our fit. We thank D.~Callos for useful discussions. We are also grateful to M. Engelhardt for providing us with the results for the worm-gear shift from Ref.~\cite{Yoon:2017qzo} and for fruitful discussions about the lattice results for $g_{1T}$. D.P.~thanks R.~Sassot for providing the tables from Ref.~\cite{DeFlorian:2019xxt} and for help with the authors' Fortran code. S.B. was supported by the U.S. Department of Energy under Contract No. DE-SC0012704. S.B. has also been supported by the U.S. Department of Energy, Office of Science, Office of Nuclear Physics and Office of Advanced Scientific Computing Research within the framework of Scientific Discovery through Advance Computing (SciDAC) award Computing the Properties of Matter with Leadership Computing Resources. This work has been supported by the National Science Foundation under grants No.~PHY-2110472 (S.B., A.M.~and G.P.), No.~PHY-1945471 (Z.K.), and No.~PHY-2011763 (D.P.). This work has also been supported by the U.S. Department of Energy, Office of Science, Office of Nuclear Physics, within the framework of the TMD Topical Collaboration. 
\end{acknowledgements}

\appendix
\section{Main fit results with replicas}
\label{app_a}
\begin{figure}[t]
\centering
\includegraphics[width=0.9\textwidth]{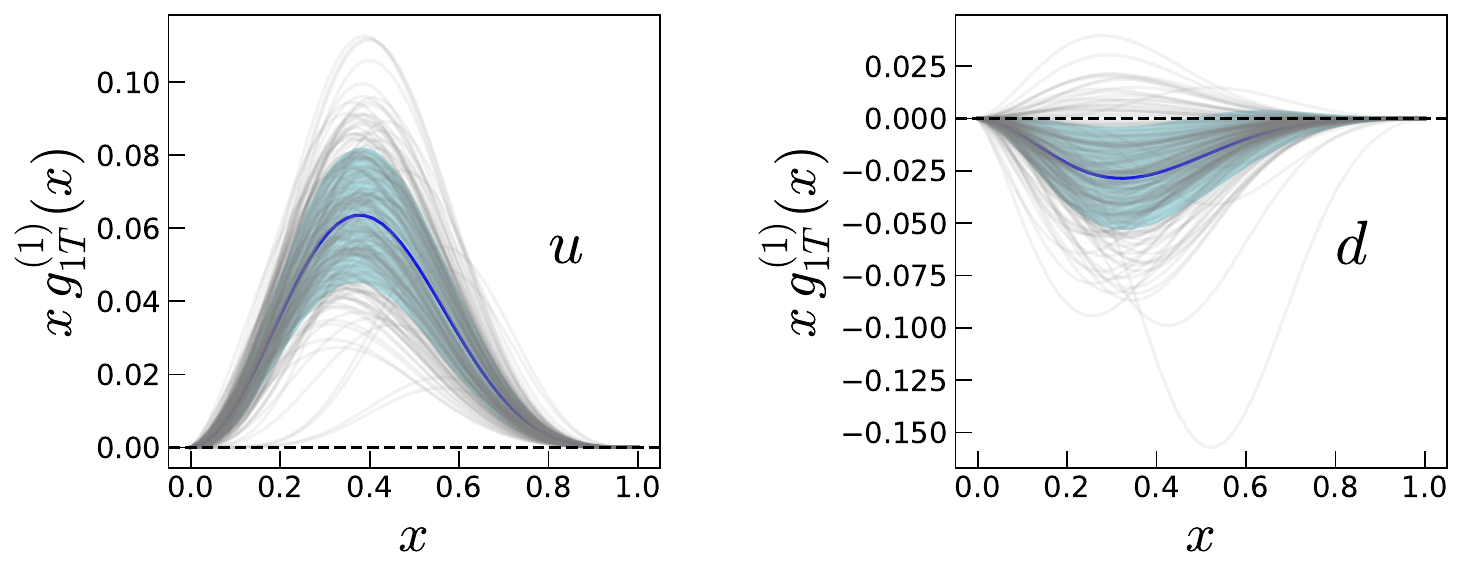}
\caption{Comparison of our error bands (blue bands) for $xg_{1T}^{(1)}(x)$ with the actual replicas (gray curves). The results have been shown at $Q^{2}$ = 4 ${\rm GeV}^2$ for up quarks (left) and down quarks (right). These results correspond to the weighted $\chi^2$ set up.}
\label{f:g1T_error_replica}
\end{figure}
In this Appendix, we show the $x g^{(1)}_{1T}(x)$ plot with the error band and all replicas to give the reader a better sense of how the functional form of $g^{(1)}_{1T}(x)$ varies as well as justify our method of error calculation.
From Fig.~\ref{f:g1T_error_replica}, we notice in particular that there are some replicas for $g^{(1)d}_{1T}(x)$ which are positive. The error bands for $g^{(1)}_{1T}(x)$ do appropriately account for $\sim 68 \%$ of the replicas as they should even though the distributions of $\chi^{2}$ themselves are not strictly Gaussian (see Fig.~\ref{f:chi2_distribution}).

\section{Additional plots from fits in the unweighted $\chi^{2}$ method}
\label{app_b}
\begin{figure}[t]
\centering
\includegraphics[width=0.9\textwidth]{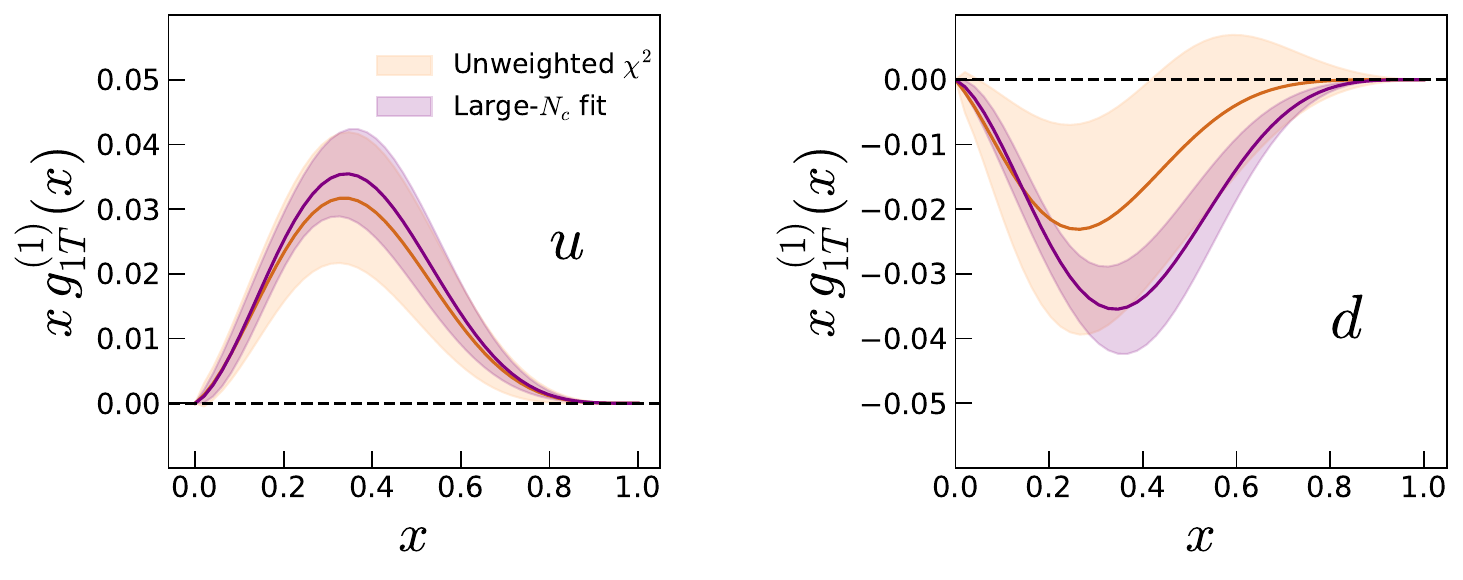}
\caption{Comparison of our main extraction of $xg_{1T}^{(1)}(x)$ at $Q^{2}$ = 4 ${\rm GeV}^2$ for up quarks (left) and down quarks (right) with the results obtained by imposing the large-$N_c$ approximation on the fit.  Both functions were obtained using the unweighted $\chi^2$ method.}
\label{f:fig_17}
\end{figure}

\begin{figure}[t]
\centering
\includegraphics[width=0.9\textwidth]{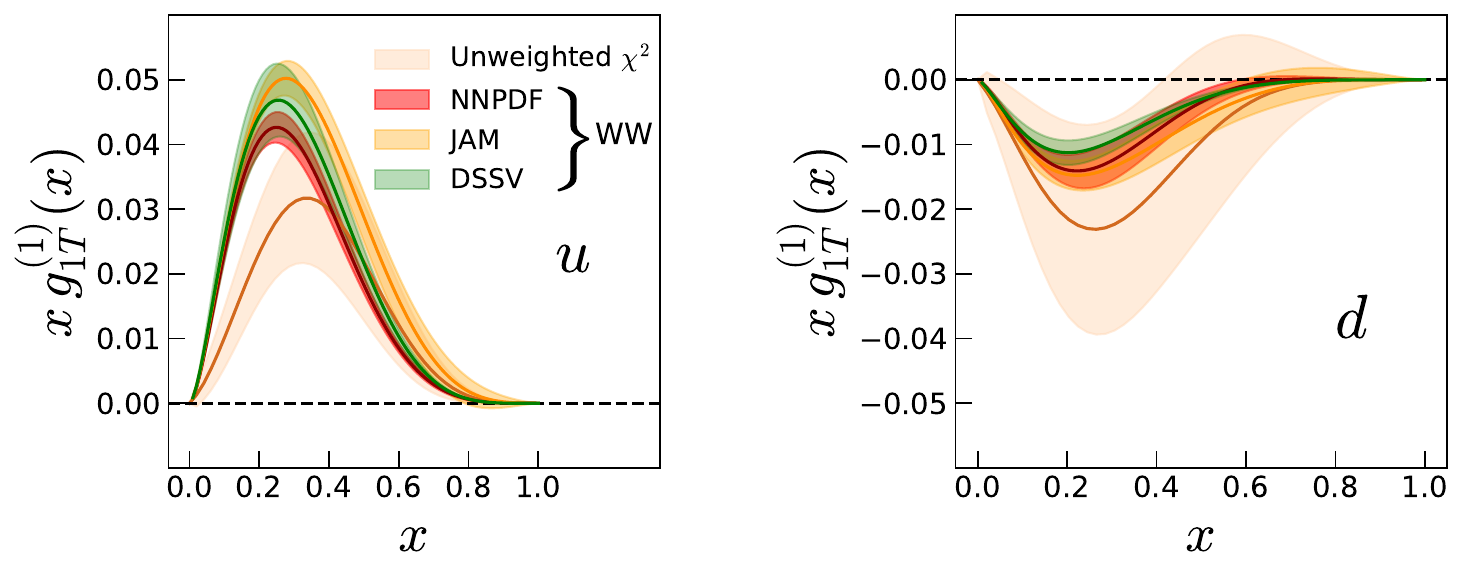}
\caption{Comparison of our unweighted $\chi^2$ extraction of $xg_{1T}^{(1)}(x)$ at $Q^{2}$ = 4 ${\rm GeV}^2$ for up quarks (left) and down quarks (right) with the results obtained from the calculation of $xg^{\rm WW}_{1T}(x)$ using Eq.~(\ref{e:WWtype}) with $g_1(x)$ taken from NNPDF~\cite{Nocera:2014gqa}, JAM~\cite{Ethier:2017zbq}, and DSSV~\cite{DeFlorian:2019xxt}.}
\label{f:fig_18}
\end{figure}

In this Appendix, we compare the fit results obtained in the unweighted $\chi^{2}$ method (Fig.~\ref{f:weighted_vs_unweighted}) with the large-$N_c$ (Fig.~\ref{f:fig_17}) and the WW-type approximations (Fig.~\ref{f:fig_18}). We notice that the large-$N_c$ results overlap rather well with these fit results. (Here the large-$N_c$ fit has been carried out using the unweighted $\chi^{2}$.) Furthermore, the results for the WW-type approximation are very interesting. There seems to be a consistent hint for a slight violation of the WW-type approximation for the up quark in the moderate-to-small-$x$ region, both in the weighted and unweighted $\chi^{2}$ approaches. However, the differences in the large-$x$ region seems to be relatively stronger in the weighted $\chi^{2}$ set up. One can speculate that such a feature is driven by the JLab data. The qualitative conclusion for the down quark remains unchanged from the main fit. 


\bibliography{references}

\end{document}